\documentclass{article}
\usepackage{amsmath}
\usepackage{amssymb}
\usepackage{graphicx}
\usepackage{subcaption}
\usepackage{authblk}
\usepackage{geometry}
\usepackage{wrapfig}
\usepackage{tikz}
\usepackage{pgfplots}
\usepackage{array}
\usepackage{float}

\pgfplotsset{compat=1.18}
\usetikzlibrary{calc,intersections,fillbetween}
\usetikzlibrary{calc}
\usetikzlibrary{positioning,arrows.meta}
\usepackage[backend=bibtex,style=numeric,sorting=none]{biblatex}
\usepackage{color}
\newcommand{\rev}[1]{{\color{black}#1}}

\title{NIMROD-to-IMAS workflow for extended-magnetohydrodynamic data interoperability}
\author[1]{Alexei Y. Pankin\thanks{Corresponding author: pankin@pppl.gov}}
\author[1]{Fatima Ebrahimi}
\author[2]{Qian Gong}
\author[3]{Jacob King}
\author[1]{Andreas Kleiner}
\author[3]{Jesus Dominguez-Palacios}
\author[2]{Norbert Podhorszki}
\author[2]{Eric Suchyta}
\affil[1]{Princeton Plasma Physics Laboratory, Princeton, NJ 08540}
\affil[2]{Oak Ridge National Laboratory, Oak Ridge, TN 37830}
\affil[3]{Fiat Lux, Lafayette, CO 80026}
\date{January 2026}
\date{\today}

\begin{document}
\maketitle

\begin{abstract}
Extended magnetohydrodynamic (MHD) simulations of tokamak plasmas regularly produce outputs in multi-dimensional, multiple-field formats; these code-specific formats make it difficult to do cross-code validation/coupling and analyze at a database scale. In this paper, a workflow that converts NIMROD code inputs and outputs to records compatible with version 4 of the ITER IMAS Data Dictionary is presented. The scope of the workflow includes preprocessing of NIMROD code inputs, conversion of hierarchical NIMROD code HDF5 dumps, COCOS-consistent treatment of the coordinate system and sign convention, and encoding finite-element poloidal meshes and toroidal Fourier components through IMAS General Grid Description. Furthermore, the workflow allows for provenance and integrity metadata to be included while providing optimal I/O operations for large array structures. An example conversion based on an NIMROD code simulation of edge harmonic oscillations performed for the DIII-D discharge 163518 [A.Y. Pankin {\em et al.}, Nuclear Fusion 60.9 (2020), p. 092004] is used to validate the conservation of essential equilibrium, profile, perturbation, and grid data in the resulting IMAS records. Finally, this implementation exposes gaps in the current IMAS framework that need to be addressed to accommodate extended MHD data, and it highlights the metadata, provenance, and governance needs of the downstream use cases in the form of dataset validation, integration, and machine learning.

\end{abstract}

\break

\begin{table*}[h!]
\label{tab}
\begin{tabular}{p{0.30\linewidth}p{0.64\linewidth}}
\hline
\textbf{Program title} & \texttt{nimrod2imas} \\

\textbf{CPC Library link to program files} & {To be added by Technical Editor} \\

\textbf{Developer's repository link} &
\url{https://github.com/PrincetonUniversity/nimrod2imas} \\

\textbf{Licensing provisions} & MIT \\

\textbf{Programming language} & Python \\

\textbf{Supplementary material} &
Source code, installation instructions, Makefile targets, a reduced NIMROD-to-IMAS test case, expected output, and validation scripts for profile and \texttt{mhd\_linear} structures. Complete test dataset is available through the associated public data repository. \\

\textbf{Nature of problem} &
The data generated by extended-MHD simulations performed with NIMROD use code-specific data formats which are not directly compatible with other code formats or \rev{formats used in experiments}. This limits the use of NIMROD simulation data in integrated modeling workflows and requires development of customized interfaces for experimental and simulation frameworks such as integrated modeling workflows, code-to-code comparison tools, data archiving, and machine learning application. \\

\textbf{Solution method} &
The code converts NIMROD input and output data to IMAS-compliant data structures by mapping NIMROD mesh, fields, equilibria, and profiles to the equivalent IMAS entries, providing IDSs for subsequent use with other IMAS-compliant programs. The functionality is illustrated using the results from a non-linear simulation of NIMROD with DIII-D discharge 163518.\\

\textbf{Additional comments including restrictions and unusual features} &
This program is intended for storing the extended-MHD NIMROD data and supports the IMAS quantities and grid representations needed for the workflow demonstrated in this paper. The emphasis is on reproducibly converting NIMROD data into IMAS format. Users are expected to verify the compatibility with the local IMAS version and with their selection of NIMROD variable subsets required for their application. A reduced example for converting NIMROD to IMAS, along with sample input and output, is provided with the program archive. As the production dataset described in the manuscript is too large, it is placed separately in the data depository.\\

\textbf{References} &
[1] C.R. Sovinec et al., J. Comput. Phys. 195 (2004) 355--386;\newline 
[2] F. Imbeaux et al., Nucl. Fusion 55 (2015) 123006; \newline
[3] A.Y. Pankin et al., Submitted to Scientific Data (2026). \\
\hline
\end{tabular}
\end{table*}

\section{Introduction}

The growing importance of fusion plasma research relies increasingly on the capacity to combine simulations of various aspects of physics with experimental results in a unified manner. Advanced simulations of tokamaks, including those involving extended magnetohydrodynamics (extended-MHD), produce large quantities of complex data that make them difficult to share or analyze across simulation tools. This is the case, for example, with the NIMROD code~\cite{Sovinec2004}, which is a high-order finite element code for solving nonlinear magnetohydrodynamics problems, including two-fluid and kinetic processes. This code has been critical in the study of various phenomena, including the generation of edge-localized modes (ELMs) or other instabilities in either traditional or spherical tokamaks. However, the structure and terminology used in the code’s output system are specific to the code itself, which may make it difficult to use or combine with other tools or modern methods of data analysis. \rev{The reality of complexity in a practical sense is that native file formats used to store simulation data are not necessarily guaranteed to be stable throughout the lifespan of the code itself. Within any individual code, such as the NIMROD code, the way in which the output quantity is structured and interpreted changes with time from one code revision to another, while the post-processing routines remain valid only for the latest formats. As a result, the analysis of simulation data stored in older formats requires either the use of the corresponding legacy code or extensive knowledge about the version and interpretation of the data. Cross-code analysis adds another level of complexity in that the same physical quantity can be stored in different units, coordinate systems, use different sign conventions, normalizations, grids, and field decompositions.}

A prevalent trend among both linear and nonlinear MHD simulation tools is their tendency to store results in formats that are native to the code and optimized for the generation workflow. This approach creates a bottleneck for cross-code verification and validation, coupling, and large-scale curation. When performing nonlinear extended-MHD simulations with NIMROD, users usually select to create a series of hierarchical HDF5 ``dump'' files. These files contain multiple levels of information, including time, grid, and fields. These dumps save large amounts of data for snapshots or for restarting the simulations. M3D-C1 also uses HDF5 as its default output format and writes the following types of information into an HDF5 file: state variables; scalar variables; equilibrium information; and time-indexed groups~\cite{Jardin2012,M3DC1_Guide}. Nonlinear edge-MHD simulations with  JOREK~\cite{Hoelzl2021} generate outputs and restart files in HDF5 format. There are existing integration efforts within the JOREK development team on the extraction of plasma-state data from the collection of HDF5 outputs at various time slices, to consolidate it into a single representation, including IMAS-oriented workflows. Many linear ideal-MHD stability tools have historically utilized ASCII equilibrium inputs and comparatively low-resource output products. For instance, the GATO manual~\cite{GATO_Manual,Bernard1981} indicates that its input equilibrium format adheres to ASCII files from EFIT~\cite{Lao1985}, reflecting the longstanding practice of text-based equilibrium exchange for eigenvalue calculations and stability assessments. 

One important milestone in the direction of data standardization in the magnetic fusion community has been the creation of the Integrated Modeling \& Analysis Suite (IMAS). \rev{IMAS offers a common Data Dictionary (DD) and data model for data storage, aiming to describe both experimental and simulation data in a similar way.} IMAS was proposed as the core for ITER future data and modeling platform, boasting a machine-independent hierarchy (the Interface Data Structures, IDSs), promoting a strict organization of plasma physics data~\cite{Imbeaux2015}\rev{, and providing a standardized data model and controlled dictionary for fusion data, supporting exchange of information between simulation codes without claiming the formal semantics of an ontology.} 

\rev{IMAS has gradually gained recognition as a universal way to represent data in an integrated approach to modeling and analysis. Among the examples are: the adoptation of IMAS-compatible data structures for supporting multi-code simulations in the OMFIT framework~\cite{Meneghini2015} and TRANSP integrated modeling code~\cite{Pankin2025}; the introduction of DINA free-boundary equilibrium code into the IMAS-driven workflow for scenario modeling, involving coupling with JINTRAC and performing the periodic stability analysis using HELENA and MISHKA~\cite{Khayrutdinov1993,Mikhailovskii1997,Koechl2018}; and equilibrium reconstruction and standardizing of the equilibrium databases provided by OMFIT/CAKE and EFIT-AI \cite{Smith2024,Lao2022,Bechtel2023}. In the case of nonlinear extended-MHD modeling, one finds the example of JOREK \cite{Hoelzl2021}, in which IMAS and the General Grid Description framework are used to describe mesh-based data on plasma state and provide a means of coupling simulation input and output \cite{Penko2019}.} The ITER Organization has recently released IMAS as open software, promoting its usage as a global standard for fusion research. This way, data will be stored in a similar fashion, making it easier for scientists to reproduce analysis and, more importantly, combine experimental and simulation data, a feature that will prove invaluable in validation analysis and upcoming ML tasks. \rev{All in all, this shows the use of IMAS as a universal interface for the tasks of equilibrium reconstruction, profile analysis, stability calculations, and nonlinear simulations, as is done here in the context of NIMROD.}

Despite this, there remain issues in the application of IMAS to new domains and physics. The IMAS data dictionary was formulated to work in the context of the tokamak and established plasma physics. There are new issues in the extended-MHD simulation, including multiple ions and their temperature, complex resistive and two-fluid models, and extended diagnostic data. At present, some of these variables lack any specific placeholder in the IMAS data dictionary. For example, there is no specific field in the IMAS data dictionary to hold the temperature and density perturbations of the second ion in the multi-ion plasma for extended MHD simulations. Thus, the incorporation of the extended-MHD simulation data into the IMAS data dictionary requires the effective management of the data dictionary gaps. This may otherwise require the researcher to introduce workarounds to the data dictionary issues, which would negatively affect data interoperability, such as assigning new data to existing fields or creating new user-defined data structures. In this work, we provide an objective assessment of the IMAS data dictionary gaps and provide solutions to these issues. Where possible, we provide solutions in the form of the existing IMAS IDS entries, such as the storage of multiple ion-temperature perturbation profiles in alternative IDS occurrences. For the data dictionary entries lacking in the IMAS data dictionary, such as the various dissipation rates or custom diagnostic data, we consider extensions to the IMAS data dictionary in line with the design guidelines of the data dictionary, to provide the potential to incorporate these extensions in the form of new IMAS data dictionary releases in the future. This is to ensure that no physics is lost in the data translation process. Thus, the researcher downloading the IMAS data dictionary-converted data from NIMROD simulation data will have simulation data from the center to the boundary as if NIMROD simulation data were directly stored in the IMAS data dictionary.

Furthermore, a significant issue is the performance. Realistic extended MHD simulations with NIMROD can produce significant amounts of two- or three-dimensional data, e.g., time series of two-dimensional sections in the poloidal plane with multiple variables or complete three-dimensional fields inside the torus. Outputting this data to a structured or unstructured formats in IMAS must be done efficiently to remain feasible. We have coded a ``data converter'', named \texttt{dump2imas}, which reads the NIMROD data in its native format and initializes the associated IMAS data structures. The data converter makes heavy use of chunk-based binary file I/O and vector operations to minimize overhead. The overhead of the data conversion step, which we have tested with a plasma simulation in the DIII-D and NSTX tokamaks, turned out to be relatively small compared to the actual simulation time. We also discuss other optimization ideas like batching file I/O operations or compressing non-critical data to keep the performance penalty small. This allows researchers to integrate the data output of the IMAS code into an ordinary NIMROD post-processing workflow of their simulations without a significant performance penalty, making the application of the framework feasible.

The NIMROD nonlinear extended MHD simulations are generally recognized as HPC-class problems based on the three-dimensional implicit time advance with associated large sparse solves repeated many times through the simulation process. For these simulations, the computational cost is affected by the nonlinear simulation process itself, but also by the requirement to generate frequent field snapshots of high fidelity for physical interpretation and synthetic diagnostics. An important case study is the validated NIMROD simulation of edge harmonic oscillation (EHO) in DIII-D quiescent H-mode discharge 163518, where comparison to experiment relied on synthetic diagnostics such as synthetic BES and mode-resolved analysis of saturated three-dimensional fluctuations~\cite{Pankin2020}. For the EHO simulation campaign, frequent state dumps were necessary to ensure temporal resolution adequate for synthetic forward modeling diagnostics and to track the evolution and saturation of low-$n$ harmonics. For the NIMROD simulation case using the same simulation parameters as the production simulation case -- spectral element mesh size of (72 x 128) with polynomial order 6 and 22 Fourier harmonics -- we find that each state dump is of order 1.3 GB size. Dumps were written every $10^{-6}$~s (or approximately every 100 time steps) towards the end of the simulation, while the nonlinear simulation process advanced through 255,000 time steps. Thus, over 2,000 dumps were generated with a corresponding storage requirement of order 3.3 TB.

Such requirements escalate very quickly with increasing physical demands and, hence, higher resolutions. For example, the wide pedestal QH mode on DIII-D is an established operating scenario that requires higher toroidal spectral content for accurate simulation of broader pedestal structures and associated 3D effects. If, for example, the toroidal representation increases from 22 to 86, and considering that the size of each dump may scale linearly with the number of toroidal modes being stored, while otherwise being similar, then each dump size increases from 1.3 GB to 5 GB, and for otherwise similar dump cadence and duration, this equates to a total storage requirement of around 12-13~TB.

In addition, ST tokamak scenarios often have additional requirements for poloidal resolution, especially considering that, for ST edge conditions, large pedestal region safety factors $q$ are often typical, and large $q$, combined with strong edge shear, means close-spaced resonant surfaces and hence strong coupling to a wide range of poloidal modes, thereby increasing poloidal resolution requirements for MHD stability and nonlinear MHD modeling.

There are multiple advantages to converting extended MHD simulation data into the IMAS format. First, there is much-improved interoperability, since once data is in the IMAS format, it can be readily input into other analysis codes and simulation tools that support it. For instance, NIMROD simulation data could be coupled into an IMAS-capable transport code or equilibrium solver, allowing multi-code iterative calculations, which is crucial for coupled simulation analyses of steady-state problems. Second, data reproducibility and data durability are improved by being in the IMAS format. Each and every IMAS dataset is self-contained, including detailed metadata about the simulation, scenario, and underlying assumptions made by the simulation code. These data are filled out by our framework, including geometry, boundary conditions, and other variables, in accordance with data standards within our research communities. This way, even after multiple years, a researcher can understand and use again NIMROD simulation data without needing to consult the original authors’ documentation or commentary. Following FAIR data guidelines~\cite{Wilkinson2016}, such detailed description and standardized representation are essential to data Findability and Reusability by other researchers and computational systems alike. 

\rev{The reverse IMAS-to-NIMROD path does not require NIMROD to ingest raw diagnostic IDSs directly. NIMROD simulations are initialized from reconstructed equilibrium and kinetic-profile information, conventionally supplied through \texttt{geqdsk} and \texttt{peqdsk} files and subsequently processed by NIMROD pre-processing tools. Existing integrated-modeling tools can already obtain these equilibrium and profile representations from IMAS data, including through OMFIT functionality. Consequently, the present workflow does not duplicate this functionality. A closed workflow can instead use the existing tools to prepare the NIMROD inputs, followed by the NIMROD-to-IMAS conversion described here for simulation outputs.}

In addition to interoperability and reusability, the standardized IMAS format also allows for a new perspective toward archival for nonlinear MHD simulations in which the concept of restart-oriented archiving may be feasible. It is possible that if the entire evolving state, the choice of model, closures, and the desired diagnostics were encoded in a single format, then such an approach would facilitate checkpoint-style recreation and improve interoperability between simulation codes. This additional perspective partially inspires the current effort. \rev{While the workflow presented here focuses on converting NIMROD data into standardized and provenance-laden IMAS entries, it does not claim restart-level completeness for all extended MHD models used by NIMROD. This limitation follows from the present IMAS DD, which does not yet provide standardized fields for all dissipation and closure quantities, multi-species perturbations, and solver-specific state needed for a comprehensive NIMROD restart.}

%We describe ‘restartability’ as a specific application area within reproducibility, where an IMAS dataset from a nonlinear MHD simulation can be used to restart a simulation or pick up where it left off if it was terminated mid-stream. With our framework, where every state variable, including evolving equilibria and profiles, is recorded in the IMAS format, it is possible to checkpoint a simulation completely, not only facilitating traditional code restarts but, in principle, even facilitating simulation continuation by other simulation codes, which is useful not only to ‘prove’ specific simulation results or outcomes by other simulation tools, testing their consistency and compatibility, but also to ‘combine’ advantages and strengths from multiple simulation tools and codes, if needed or applicable.

Finally, our aim is in line with future AI/ML tasks for fusion. There is an increasing appreciation that data readiness is a bottleneck for applying modern ML to scientific research. The fusion community is not an exception; while there is considerable interest in applying ML to disruption prediction, surrogate modeling of plasma profiles, and fast design of experiments, such analyses are commonly impeded by data fragmentation and inconsistencies. In this respect, by harmonizing simulation data from extended-MHD models to a common format and ensuring completeness and structure, we make these data more amenable to systematic feature extraction and downstream ML analysis. For example, it is possible to train an ML neural network on an archive of NIMROD simulations in the IMAS format to replicate specific plasma phenomena. Because IMAS has an inherently hierarchical, labeled data structure, an ML process can automatically access relevant data (e.g., temperature distributions, current density) for multiple discharges or simulations without having to manually process data from multiple files. Most importantly, data standardization makes it feasible to perform multi-modal ML; for instance, an ML process can analyze data from DIII-D or ITER experiments and NIMROD simulation data using exactly the same data structure, benefiting from both sources of data -- something that cannot be done if results were to be expressed in code-specific data formats. Thus, our work will help make fusion data generally accessible to AI/ML tools. There have been recent community reports emphasizing that data models such as IMAS have an important future for applying AI to fusion research, strongly encouraging that such models should be used to represent data from future research, such that data from existing research will have to conform to this format. We propose to tackle this challenge of storing data from extended-MHD models in an IMAS format and make this data openly available to help provide a basis for future AI/ML tools for fusion energy research. In short, this paper presents the design and implementation of a data framework for the NIMROD extended MHD code based on the IMAS framework. 

The main objectives of this research are threefold. Firstly, the workflow is presented to convert the simulation results generated by NIMROD MHD codes into datasets compatible with the IMAS format. Secondly, through the realization of this workflow, we can recognize the shortcomings and needs in terms of expanding IMAS data models when dealing with extended MHD simulations. Lastly, the research highlights some of the metadata, provenance, performance, and governance considerations required for building reusable campaigns that facilitate downstream validation, model integration, and future machine learning tasks. A sample conversion using a DIII-D QH simulation scenario is used to illustrate the workflow and validate the records.

A dataset-centered description of the deposited example IMAS record is provided separately as a Data Descriptor~\cite{pankin2026sd}; the present paper focuses on the conversion workflow, mapping choices, validation strategy, performance considerations, and implications for IMAS schema development.

This paper presents a workflow for converting NIMROD magnetohydrodynamic simulation data into IMAS-compliant records. Section 2 describes the IMAS data model, the requirements posed by NIMROD outputs, and the implementation of the conversion workflow, including coordinate-convention handling, mesh encoding, performance considerations, and provenance capture. Section 3 presents results from an example conversion of a DIII-D quiescent H-mode simulation and summarizes validation of the resulting IMAS records. Section 4 discusses interoperability, reuse, current limitations of the workflow, and possible extensions for database-scale and machine-learning applications.

\section{NIMROD-to-IMAS conversion workflow\label{sec:sec3}}

The fusion community has increasingly used the IMAS schema to provide a common representation of data and to enable interoperability among experiments, theory, and multi-physics simulations. The IMAS schema is built upon a machine-generic data model called the IDSs and aims to provide a common representation of experimental and simulated data in terms of their semantics and units and their coordinates~\cite{Imbeaux2015}. The common representation of data provided by IMAS has reduced the need for custom file formats and custom data converters and has enabled end-to-end workflows in which equilibrium reconstruction and stability analyses can communicate their results to nonlinear MHD simulations.

\subsection{IMAS and NIMROD extended-MHD requirements}

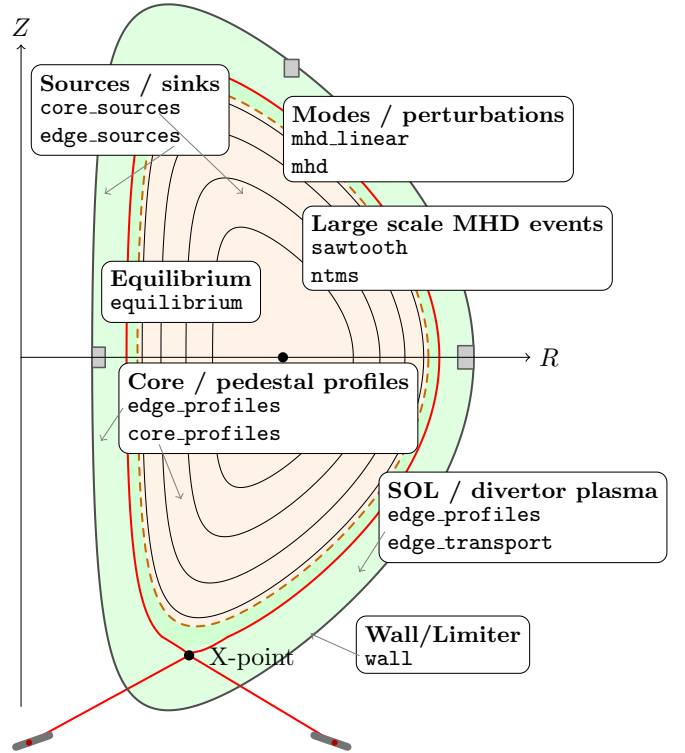
\begin{wrapfigure}{r}{0.48\textwidth}
  \centering
  \resizebox{0.41\textwidth}{!}{
    \begin{tikzpicture}[scale=1.0, line cap=round, line join=round]
    \path[use as bounding box] (0.4,-5.3) rectangle (7.2,4.1);
% Equilibrium parameters
  \def\R0{3.6}
  \def\Z0{0.0}
  \def\a{2.15}
  \def\kappa{1.85}
  \def\delta{0.6}
  \pgfmathsetmacro{\asinDel}{asin(\delta)} % degrees

  % Scales
  \def\sPed{0.93}   % pedestal surface (inside LCFS)
  \def\sWall{1.22}  % wall scale (>1). Try 1.18--1.30

  % --- Pedestal boundary (closed) ---
  \path[name path=PED]
    plot[smooth, samples=260, domain=0:360]
    ({\R0 + (\sPed*\a)*cos(\x + \asinDel*sin(\x))},
     {\Z0 + (\sPed*\kappa*\a)*sin(\x)}) -- cycle;

  % --- LCFS-shaped boundary (closed; used for fills only) ---
  \path[name path=LCFS_SHAPE]
    plot[smooth, samples=360, domain=0:360]
    ({\R0 + \a*cos(\x + \asinDel*sin(\x))},
     {\Z0 + \kappa*\a*sin(\x)}) -- cycle;

  % --- Wall (closed; aligned with LCFS) ---
  \path[name path=WALL]
    plot[smooth, samples=360, domain=0:360]
    ({\R0 + (\sWall*\a)*cos(\x + \asinDel*sin(\x))},
     {\Z0 + (\sWall*\kappa*\a)*sin(\x)}) -- cycle;

  % Wall outline (draw later)
\def\WallDraw{%
  \draw[black!70, thick]
    plot[smooth, samples=360, domain=0:360]
    ({\R0 + (\sWall*\a)*cos(\x + \asinDel*sin(\x))},
     {\Z0 + (\sWall*\kappa*\a)*sin(\x)});
}

  % --- Limiters (on wall at poloidal angles) ---
  \def\thLimA{0}
  \def\thLimB{55}
  \def\thLimC{180}

  \coordinate (WLA) at ({\R0 + (\sWall*\a)*cos(\thLimA + \asinDel*sin(\thLimA))},
                        {\Z0 + (\sWall*\kappa*\a)*sin(\thLimA)});
  \coordinate (WLB) at ({\R0 + (\sWall*\a)*cos(\thLimB + \asinDel*sin(\thLimB))},
                        {\Z0 + (\sWall*\kappa*\a)*sin(\thLimB)});
  \coordinate (WLC) at ({\R0 + (\sWall*\a)*cos(\thLimC + \asinDel*sin(\thLimC))},
                        {\Z0 + (\sWall*\kappa*\a)*sin(\thLimC)});

  \def\limDX{0.22}
  \def\limDZ{0.16}

  \def\LimiterDraw{%
    \begin{scope}
      \fill[black!20] (WLA) ++(-\limDX,-\limDZ) rectangle ++(\limDX,2*\limDZ);
      \draw[black!65, line width=0.6pt] (WLA) ++(-\limDX,-\limDZ) rectangle ++(\limDX,2*\limDZ);

      \fill[black!20] (WLB) ++(-0.20,-0.12) rectangle ++(0.20,0.24);
      \draw[black!65, line width=0.6pt] (WLB) ++(-0.20,-0.12) rectangle ++(0.20,0.24);

      \fill[black!20] (WLC) ++(0.00,-0.14) rectangle ++(0.18,0.28);
      \draw[black!65, line width=0.6pt] (WLC) ++(0.00,-0.14) rectangle ++(0.18,0.28);
    \end{scope}
  }

  % --------------------------
  % Shading (draw FIRST)
  % --------------------------

  % Core: inside PED
  \path[fill=orange!10]
    plot[smooth, samples=260, domain=0:360]
    ({\R0 + (\sPed*\a)*cos(\x + \asinDel*sin(\x))},
     {\Z0 + (\sPed*\kappa*\a)*sin(\x)}) -- cycle;

  % Pedestal band (between PED and LCFS_SHAPE)
  \path[fill=green!18, even odd rule]
    plot[smooth, samples=360, domain=0:360]
    ({\R0 + \a*cos(\x + \asinDel*sin(\x))},
     {\Z0 + \kappa*\a*sin(\x)}) -- cycle
    plot[smooth, samples=260, domain=0:360]
    ({\R0 + (\sPed*\a)*cos(\x + \asinDel*sin(\x))},
     {\Z0 + (\sPed*\kappa*\a)*sin(\x)}) -- cycle;

  % SOL region (between LCFS_SHAPE and WALL)
  \path[fill=green!12, even odd rule]
    plot[smooth, samples=360, domain=0:360]
    ({\R0 + (\sWall*\a)*cos(\x + \asinDel*sin(\x))},
     {\Z0 + (\sWall*\kappa*\a)*sin(\x)}) -- cycle
    plot[smooth, samples=360, domain=0:360]
    ({\R0 + \a*cos(\x + \asinDel*sin(\x))},
     {\Z0 + \kappa*\a*sin(\x)}) -- cycle;

  % --------------------------
  % Flux surfaces
  % --------------------------
  \foreach \s in {0.45,0.62,0.78,0.90} {
    \draw[black]
      plot[smooth, samples=260, domain=0:360]
      ({\R0 + (\s*\a)*cos(\x + \asinDel*sin(\x))},
       {\Z0 + (\s*\kappa*\a)*sin(\x)});
  }

  % --------------------------
  % Separatrix with X-point
  % --------------------------
  \def\thL{255}
  \def\thR{285}

  \coordinate (SepL) at ({\R0 + \a*cos(\thL + \asinDel*sin(\thL))},
                         {\Z0 + \kappa*\a*sin(\thL)});
  \coordinate (SepR) at ({\R0 + \a*cos(\thR + \asinDel*sin(\thR))},
                         {\Z0 + \kappa*\a*sin(\thR)});

  \coordinate (Xpt) at ({\R0 - 0.6*\a},{\Z0 - 1.03*\kappa*\a});

  % LCFS (excluding [thL, thR])
  \draw[red, thick]
    plot[smooth, samples=360, domain=\thR:360]
    ({\R0 + \a*cos(\x + \asinDel*sin(\x))},
     {\Z0 + \kappa*\a*sin(\x)})
    --
    plot[smooth, samples=360, domain=0:\thL]
    ({\R0 + \a*cos(\x + \asinDel*sin(\x))},
     {\Z0 + \kappa*\a*sin(\x)});

  % Cusp branches into X-point
  \draw[red, thick]
    (SepL)
    .. controls ($(SepL)!1.2!(Xpt) + (-0.1,0.10)$) and ($(Xpt) + (-0.05,0.05)$)
    .. (Xpt);

  \draw[red, thick]
    (Xpt)
    .. controls ($(Xpt) + (0.10,0.1)$) and ($(SepR)!0.95!(Xpt) + (0.05,-0.05)$)
    .. (SepR);

  % --------------------------
  % Divertor
  % --------------------------
  \coordinate (TgtL) at ($(Xpt)+(-2.2,-1.2)$);
  \coordinate (TgtR) at ($(Xpt)+( 2.0,-1.2)$);

  \draw[red, thick] (Xpt) -- (TgtL);
  \draw[red, thick] (Xpt) -- (TgtR);

  \draw[black!55, line width=3.0pt]
    ($(TgtL)+(-0.18,-0.06)$) -- ($(TgtL)+(0.28,0.10)$);
  \draw[black!55, line width=3.0pt]
    ($(TgtR)+(-0.28,0.10)$) -- ($(TgtR)+(0.18,-0.06)$);

  \fill[red!70!black] (TgtL) circle (1.2pt);
  \fill[red!70!black] (TgtR) circle (1.2pt);

  % Pedestal boundary (dashed)
  \draw[orange!80!black, thick, dashed]
    plot[smooth, samples=260, domain=0:360]
    ({\R0 + (\sPed*\a)*cos(\x + \asinDel*sin(\x))},
     {\Z0 + (\sPed*\kappa*\a)*sin(\x)});

  % --------------------------
  % Wall outline + limiters (ON TOP)
  % --------------------------
  \WallDraw
  \LimiterDraw

  % --------------------------
  % Markers 
  % --------------------------
  \fill (\R0,\Z0) circle (2pt);
  \fill (Xpt) circle (2pt);
  \node[anchor=west] at ($(Xpt)+(0.15,-0.05)$) {X-point};

  % --------------------------
  % IDS annotation boxes 
  % --------------------------
  % NOTE: IDS naming is schematic; adjust to your IMAS version/conventions.
  \node[draw, rounded corners, fill=white, align=left, font=\small]
    at (2.2,0.9) {\textbf{Equilibrium}\\[-2pt]\texttt{equilibrium}};
  \node[draw, rounded corners, fill=white, align=left, font=\small]
    at (3.4,-0.7) {\textbf{Core / pedestal profiles}\\[-2pt]\texttt{edge\_profiles}\\\texttt{core\_profiles}};
  \node[draw, rounded corners, fill=white, align=left, font=\small]
    at (5.6,3) {\textbf{Modes / perturbations}\\[-2pt]\texttt{mhd\_linear}\\\texttt{mhd}};
  \node[draw, rounded corners, fill=white, align=left, font=\small]
    at (6.0,1.5) {\textbf{Large scale MHD events}\\[-2pt]\texttt{sawtooth}\\\texttt{ntms}};   
  \node[draw, rounded corners, fill=white, align=left, font=\small]
    at (6.9,-2.2) {\textbf{SOL / divertor plasma}\\[-2pt]\texttt{edge\_profiles}\\\texttt{edge\_transport}};
  \node[draw, rounded corners, fill=white, align=left, font=\small]
    at (5.8,-3.95) {\textbf{Wall/Limiter}\\[-2pt]\texttt{wall}};

  % Sources / sinks
  \node[draw, rounded corners, fill=white, align=left, font=\small]
    at (1.5,3.4) {\textbf{Sources / sinks}\\[-2pt]\texttt{core\_sources}\\\texttt{edge\_sources}};

  % Arrows 
  \draw[->, black!50] (1.9,3.35) -- (3.05,2.25);
  \draw[->, black!50] (2.1,2.9) -- (1.15,2.25);
  %\draw[->, black!50] (5.15,2.65) -- (5.05,2.25);
  %\draw[->, black!50] (5.15,2.65) -- (4.05,2.25);
  %\draw[->, black!50] (5.25,1.55) -- (4.95,1.05);
  %\draw[->, black!50] (5.35,0.15) -- (4.85,0.10);
  \draw[->, black!50] (1.4,-0.7) -- (1.1,-1.15);
  \draw[->, black!50] (1.9,-1.2) -- (2.2,-1.95);
  \draw[->, black!50] (5,-2.4) -- (4.65,-2.95);
  \draw[->, black!50] (4.65,-4.1) -- (3.98,-3.8);
  %\draw[->, black!50] (5.3,-4.2) -- (4.5,-5.3);

  % Arrow from sources box to edge/SOL region

  % --------------------------
  % Axes LAST
  % --------------------------
  \draw[->] (0,0) -- (7.0,0) node[right] {$R$};
  \draw[->] (0,-4.8) -- (0,4.3) node[above] {$Z$};
    \end{tikzpicture}
    }
    \caption{Schematic illustration of the plasma regions, magnetic topology, and corresponding IMAS data structures used within this research. }
    \vspace{-1.2\baselineskip}\label{fig:equilibrium}
\end{wrapfigure}
In addition to facilitating interoperability, a standardized data model has become ever more critical to a data-driven {\em modus operandi} in database curation, post-processing on a larger scale, and building analysis-ready datasets amenable to statistical inference and machine learning analysis. Examples in this direction include workflow frameworks like OMFIT, which demonstrate this trend by offering a seamless integration capability to process input files containing equilibria and profiles, multiple code execution, and result extraction in a standardized form suitable for analysis~\cite{Meneghini2015}. Thus, IMAS provides a real means to enable a FAIR-like approach to fusion data assets by making larger simulation and experimental results more findable, comparable, and reusable in a broader sense~\cite{Imbeaux2015}.

Equilibrium data are a core input to almost all modelling and analysis activities in tokamak science, from transport analysis to linear stability studies and nonlinear MHD simulations. Schematic representation of tokamak equilibrium with related IDSs is shown in Fig.~\ref{fig:equilibrium}. In this figure, the core, pedestal, scrape-off layer (SOL), and divertor regions are defined with respect to the last closed flux surface (LCFS) and the X-point of a lower single-null magnetic configuration. Core and pedestal profiles are represented within the \texttt{core\_profiles} and \texttt{edge\_profiles} IDSs, while magnetic equilibrium and separatrix geometry are represented within the \texttt{equilibrium} IDS. Linear and nonlinear perturbations within the edge region due to ELMs and ELM-free scenarios are represented within the \texttt{mhd\_linear} IDS. For the NIMROD code, the most concise and accurate representation of the non-axisymmetric information contained in a simulation (whether calculated from linear eigenmode solutions or nonlinear evolution) is given by the set of toroidal Fourier components of each field. Thus, we have stored the complex Fourier harmonics of the perturbation fields in the IMAS data structure \texttt{mhd\_linear} IDS both for linear and nonlinear simulations. This is the most appropriate location for mode-resolved information and is consistent with the NIMROD code representation and output. By using the equilibrium and $n = 0$ information in combination with the stored harmonics, the full 3D information can be reconstructed at any toroidal angle.

SOL and divertor plasma and corresponding material surface fluxes are represented with respect to edge-region IDSs such as \texttt{edge\_profiles} and \texttt{edge\_transport}, while particle and energy sources and sinks such as auxiliary heating, fueling, and radiation losses are represented within \texttt{core\_sources} and \texttt{edge\_sources} IDSs. This provides a clear and traceable mapping between NIMROD simulations, experimental inputs, and standardized IMAS databases to enable machine learning studies on ELM and ELM-free scenarios.
  
Equilibrium analysis tools are the most natural application area for integration with IMAS. The EFIT suite of equilibrium codes and similar tools solves the Grad-Shafranov equations with magnetic and auxiliary diagnostic constraints to obtain a poloidal magnetic flux, plasma boundary, and derived 1-D profiles entered into physics tools~\cite{Lao1985}. The equilibrium component of the IMAS will support a standardized storage format for this data, including 2-D flux arrays, boundary representation, and values on flux surfaces, making them easily reusable in a modelling chain~\cite{Imbeaux2015}.

\rev{Existing IMAS-based integrated-modeling workflows provide relevant precedents without requiring a detailed review of each application. In DINA+JINTRAC, equilibrium and profile information is exchanged through IMAS during time-dependent scenario modeling~\cite{Koechl2018}. For nonlinear extended-MHD, JOREK is the closest analogue to the present work: JOREK integration uses IMAS both for archiving mesh-based MHD outputs and, in scenario workflows, for supplying machine, equilibrium, and profile information used to initialize simulations~\cite{Penko2019,Massanova2025}. These examples demonstrate the bidirectional role of IMAS as both an output standard and an input/coupling interface.}

\rev{The software package discussed here handles the conversion of data from NIMROD to IMAS. In the reverse case, NIMROD does not usually import diagnostic IDSs in raw form. Initialization of experiments makes use of the reconstructed magnetic equilibrium and kinetic profile data, which is traditionally obtained using g-EQDSK and p-EQDSK files that are then handled by FGNIMEQ and the NIMROD pre-processing utilities. Current integrated modeling systems are able to provide means of obtaining such representations from the IMAS data, including via OMFIT. Therefore, a workflow can be created wherein the traditional IMAS/OMFIT tools are used to produce the g-EQDSK and p-EQDSK data required for FGNIMEQ/NIMROD, followed by the IMAS-NIMROD conversion presented here for the output simulation data.}

\subsection{IMAS Data Model and Interface Data Structures (IDS)}

In this section, we summarize the specific IDSs used in this work and the roles they play in representing equilibrium, profiles, and MHD data. %IMAS is designed to offer a standardized, computer-programming-language-independent data model for fusion research, aiming at promoting code integration, data sharing, and reproducibility across different devices. The application of a shared ontology of IDS will allow different simulation codes and analysis tools for experiments to communicate with each other without the use of data converter applications. This will enable modular, plug-and-play integration of physics codes, thereby promoting an approach that is aligned with FAIR data principles. This is an important requirement for promoting machine learning applications in fusion. For instance, tokamak-based equilibrium reconstruction data will be able to be shared in an agreed-upon form using IMAS and immediately exploited for ML applications without requiring file translation.
\rev{Out of the many standardized IDSs in the IMAS Data Dictionary (DD) v4 release, some are highly relevant to magnetohydrodynamic codes such as NIMROD and other extended-MHD codes~\cite{IMASDDv411}.} These IDSs define a common platform for the exchange of equilibrium and plasma profile data, and for storing the results from an MHD simulation:

\begin{itemize}
\item \texttt{equilibrium} is used to describe the axisymmetric magnetic equilibrium as well as profiles. This IDS covers the geometry of flux surfaces, including intersections with the major magnetic axis, separatrix, and wall; it includes two-dimensional profiles of plasma poloidal flux, $\psi(R, Z)$; profiles of toroidal flux, $\Phi$; plasma profiles including plasma pressure and current density, and some derived profiles, such as safety factor $q$ as a function of flux function, $q(\psi)$.

\item \texttt{core\_profiles} and \texttt{edge\_profiles} hold one-dimensional profiles of plasma properties as a function of the poloidal flux coordinate or other flux coordinate for the core or the SOL regions. In these profiles, values for electron or ion density, temperature, plasma rotations, or effective charge ($Z_{\rm eff}$) are specified for the core or for the SOL. It is worth noting that the division into core and SOL profiles is based on historical development practices for core transport or SOL plasma. In practice, however, \rev{coupling separately represented core and edge profiles can require careful treatment of the interface to maintain continuity of plasma quantities, gradients, and fluxes, particularly when the two regions use different grids or transport models~\cite{Hakim2012}}. Dividing into core and SOL profiles is still useful for codes that are specialized for particular regions (e.g., a core transport solver or a SOL plasma solver) to add their calculated values into a unified IMAS database. In the context of IMAS integration, NIMROD can read \texttt{core\_profiles} (for initial pressure, current, and rotations in the confined plasma) and \texttt{edge\_profiles} (for boundary conditions in the SOL, if desired) in a standardized fashion. However, separating these profiles from the NIMROD output is not practical or efficient.

\item \texttt{mhd} and \texttt{mhd\_linear} are specific to storing data for MHD simulation, providing a way to discriminate between nonlinear time-dependent and linear stability data. The \texttt{mhd} IDS is tailored for storing data for nonlinear MHD simulations, e.g., NIMROD or JOREK, over time. The IDS has room for a sequence of time slices, which have two- or three-dimensional fields (e.g., perturbations in magnetic fields, flow velocity, pressure, and electric current), and these fields are stored on the grid defined in the same IDS. To allow for arbitrary mesh data and high-dimensional data, a General Grid Description (GGD) data structure has been developed. The data structure provides a machine-independent description for a structured or unstructured mesh, which is two- or three-dimensional, and connects all stored fields to this mesh.
\end{itemize}

Basic IDSs relevant to the extended MHD simulations are shown in Fig.~\ref{fig:equilibrium}. Besides the physics-facing IDSs displayed in Fig.~\ref{fig:equilibrium}, there are several ``meta'' IDSs of importance for database-scale curation, reproducibility, and ML readiness on which our IMAS representation depends. The \texttt{summary} IDS offers a human- and machine-readable synopsis of the dataset in compact form (device, discharge/run identifiers, time range, key physics tags such as ELMy vs.\ ELM-free/QH-mode, and pointers to the primary IDS occurrences). Fast indexing and filtering of large collections without running through the whole equilibrium/profile/MHD trees is what this IDS is designed to provide support for; it becomes important in the database and ML workflows discussed in this paper. 
%The \texttt{dataset\_description} IDS is used to store the structured provenance and context information required to make the datasets self-describing. This includes code name and version, run identifiers, input deck references or hashes, coordinate and normalization conventions, and other metadata that are important for interpreting the datasets. This is consistent with the emphasis presented in the paper that the IMAS datasets are self-contained and traceable. This will help with verification/validation and interoperability between codes without the need for external bookkeeping. 
The \texttt{dataset\_fair} IDS is used to store the metadata required for the sharing and reuse of the datasets. This includes access/licensing information, persistent identifiers where available, ownership information, keywords, and integrity descriptors. This is important for the construction of ML-ready datasets where consistent metadata and unambiguous dataset lineage are required. There is also a \texttt{temporary} IDS that is used for storing startup and configuration parameters that are not currently defined clearly elsewhere in the IMAS Data Dictionary but are required for full interpretability of NIMROD simulations. This is a holding area that will be used to store the equivalent information that will be moved to the standard IDS fields as the schema is updated.

IMAS is still evolving and it has some limitations. One such limitation is related to the metadata overhead and the complexity of the model. As a result of the high level of generality that has been built into each individual IDS to accommodate a wide range of use cases, the number of fields that can be present within the model can be considerable. Furthermore, a large number of the fields can be irrelevant to the specific code or use case. As a consequence, the overall file can become quite large. Building a minimum working dataset can require populating a deep hierarchy of sub-structures. \rev{The Data Dictionary v4 redesign addressed a number of earlier ambiguities and non-backward-compatible definitions, with migration information maintained as part of the DD documentation~\cite{IMASDDv411}.} Another limitation has been the overall performance with large-scale input/output operations. For example, the ability to store high-resolution 3-D simulations with many slices within the IMAS model can prove to be a challenge. The verbose nature of the model and the overhead that the Access Layer has on many small pieces of information can prove to be a bottleneck. \rev{To mitigate this challenge, the IMAS development effort has introduced improved storage/backend capabilities, including compression support, while large-scale use cases can still require additional optimization.} However, large-scale users have realized that there can be a need for optimization. \rev{For example, writing high-resolution 3-D results at every stored NIMROD time slice can create very large data volumes and I/O costs.} There can be a need to use a form of down-sampling.
\rev{There is also a limitation related to extended-MHD application is the presence of specific physics variables within the MHD-related IDSs. Although support for multi-species has been achieved elsewhere in the IMAS Data Dictionary, such as in the species-indexed ion arrays in \texttt{edge\_profiles}, the current \texttt{mhd\_linear} IDS does not include a generalized species-indexed version of mode-resolved perturbations. Moreover, some extended-MHD variables like dissipative and closure coefficients that vary with position are not yet included in standardized places appropriate for NIMROD mapping. These limitations can be considered as lacking representations for specific extended-MHD variables, but not multi-species capabilities in IMAS.}
%Another issue that has been realized with the model has been the physics content. The IMAS DD has been built with a focus on the ITER needs with a focus on a deuterium plasma with a single fluid MHD model, some advanced physics capabilities are not fully accounted for in the schema. Some examples include: multi-species plasmas (e.g., D-T mixes and the addition of other species beyond a simple effective $Z\_{\rm eff}$), as well as higher-order fluid descriptions or non-Maxwellian distribution functions. Furthermore, the level of support within the structure for the description of anisotropic pressure tensors and higher moments thereof can be limited, often necessitating the addition of user-defined structures to combine this information. 
These limitations have been recognized as issues within user forums and discussions at a variety of user conferences and working groups~\cite{Meneghini2018,Meneghini2019,Bechtel2023}. The DD structure is continually evolving as the user base actively requests the addition of new structures to the DD structure to accommodate these additional requirements.

Nevertheless, IMAS has been shown to be an important and successful tool for tokamak scenario analysis and coupling codes, such as the combination of the DINA and JINTRAC codes through IDSs, and for experiment analysis, although still a work in progress. \rev{Feedback and suggestions for improvement regarding the data model’s metadata, I/O operations, and the overall completeness of the physics content can help guide further development.} Lessons learned regarding the integration and interaction between codes, such as the JOREK and the extensive database used in the EFIT-AI project, will help guide the NIMROD and IMAS integration and the mapping of the NIMROD extended-MHD capabilities onto the IMAS schema and will help address any shortcomings in the data model, for instance, the treatment of multiple ion species and two-fluid models.

\subsection{NIMROD Outputs and Requirements for Standardization}

The extended-MHD initial-value NIMROD code is used for studying a variety of plasma problems~\cite{Sovinec2004}. Among these plasma physics problems, it is also used for studies of edge and pedestal instabilities relevant to transient transport and edge-localized-mode (ELM) physics~\cite{Pankin2006}.  Typically, NIMROD uses the Braginski formulation of resistive magnetohydrodynamics and includes the possibility of including multi-fluid effects, \rev{which may play an important role in determining how non-ideal effects affect the stability and nonlinear saturation of instabilities in real-world tokamak configurations}~\cite{Pankin2025a}. NIMROD uses high-order finite elements in the poloidal direction, and employs a Fourier method in the toroidal direction~\cite{Sovinec2004}. This approach has been successful in resolving the complex structure of the toroidal modes important for the stability of the pedestal, providing an opportunity to compare with experimental diagnostics~\cite{Pankin2020}. 

The NIMROD code has been used in a number of detailed studies of edge stability and the nonlinear responses to extended-MHD effects~\cite{Pankin2006,King2017,Pankin2020,Pankin2025,Dominguez2026}. For example, studies on pedestal dynamics have included how physical effects such as collisionality, finite Larmor radius (FLR), and parallel closure modeling contribute to the growth rate and classification of edge modes. The studies have also shown that the inclusion/exclusion of certain effects may greatly change observed stability trends during parameter studies. A similar series of studies has shown a relationship between extended-MHD simulations of QH-mode and the observed low-$n$ edge activity~\cite{King2017,Pankin2020}. For the DIII-D tokamak, the NIMROD modeling has been used for analysis of low-torque discharges during QH-mode in order to validate mode shapes and behavior against empirical measurements, providing insight into edge harmonic oscillations and their relationship with equilibrium and edge transport. These studies expand our understanding of the multi-dimensional nature of physics effects, including the H-mode pedestal size and shape, pedestal gradient, rotational speed, impurity content/type, and closure model. The studies show that a systematic approach using a database is far more effective than isolated simulations of any of the attributes noted above. The expanded scope of this research also includes the consideration of the effect of multi-ion species collisionality on the stability of pedestal electron-directed modes within the pedestal for QH-modes in wide pedestal configurations~\cite{Dominguez2026}. Results from linear simulations performed with DIII-D wide pedestal QH-modes show that while two-fluid and ion gyroviscous dynamics can destabilize the electron-directed mode instability at the pedestal, collisionally-averaged multi-ion species can reduce the growth rate and shift the stability boundary related to the presence of an impurity. Additionally, ELM modeling and control research have developed alternative reduced or coupled modeling approaches for ELM dynamics and control, including strategies involving stochastic magnetic fluctuations within pedestal models to examine ELM suppression and altered bursting behavior. These findings advocate for a transparent and trackable methodology for presenting equilibrium and profile inputs, specifying extended-MHD model options for simulations, and detailing resultant mode changes and observable outcomes, particularly in the context of establishing reproducible workflows for multi-code integration, verification, or extensive database operations. 

An important step in developing machine learning solutions to examine ELM effects and ELM-free scenarios is the creation of a comprehensive database to encompass both ELMy and ELM-free operational domains, which goes beyond just unstable pedestals. Wide-pedestal and enhanced-pedestal H-modes are examples of ELM-free phases in spherical tokamaks and therefore have very different stability characteristics and physical interactions that are relevant for machine learning models that differentiate between highly unstable, borderline, and stable pedestal conditions amidst real uncertainty and control mechanisms. An effective database should maintain self-consistent equilibrium and profile data, uniformly record mode changes and growth-rate indicators, and standardize coordinates, units, and nomenclature to facilitate scalable data utilization. Hence, a standardized data model proves beneficial: it provides a consistent framework that supports database-scale processing, consistent features like pedestal gradients and mode growth rates, and learning that synthesizes experimental reconstructions with NIMROD simulations without requiring specialized translation layers. Ultimately, such a database aids in supervised modeling of stability margins and scenario optimization loops aimed at avoiding ELMs while preserving performance, with a focus on reproducibility and interoperability.

\begin{figure}
    \centering
    \includegraphics[width=1\linewidth]{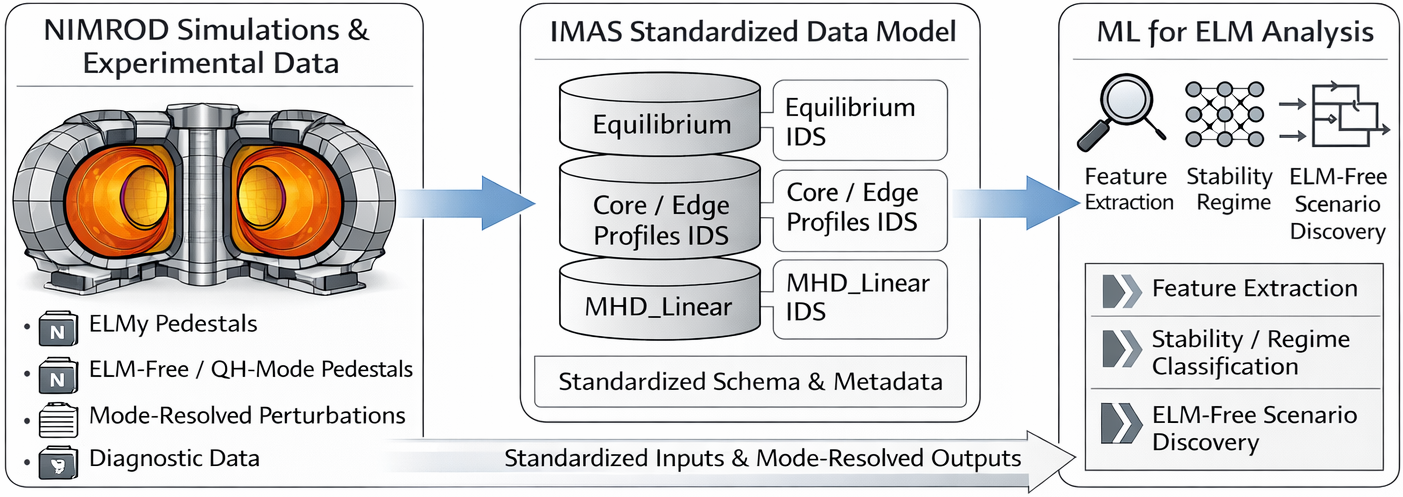}
    \caption{Schematic overview of the data flow from NIMROD simulations and experimental measurements to machine–learning–based analysis of edge-localized-mode (ELM) physics. On the left, NIMROD extended-MHD simulations and experimental inputs provide equilibrium reconstructions, core and edge profiles, and mode-resolved perturbations for both ELMy and ELM-free (e.g., QH-mode) pedestal regimes. These heterogeneous inputs are mapped into the IMAS standardized data model (center), where equilibrium, core/edge profiles, and linear MHD perturbations are stored in well-defined IDSs with consistent metadata and schema. The standardized IMAS representation enables reproducible feature extraction and large-scale database construction, which in turn supports machine-learning workflows (right) for stability classification, identification of ELM-free operating regimes, and discovery of candidate scenarios for ELM avoidance and control.}
    \label{fig:nimrod_schematic}
\end{figure}

Each NIMROD output dump file offers a snapshot of the entire state of a plasma at a specified point in time. These dumps include a full range of information, from all field quantities, such as magnetic field components, flow velocity, density, and pressure profiles, to important derived quantities, including quantities such as current density and safety factor profiles, to even model coefficient values that vary in space, such as resistivity or thermal conductivity. The snapshot model of NIMROD output means that a given output file contains everything required to restart a simulation from a specified point in time. As a matter of standardization, this means that a given format must support a full state representation of a plasma, from field quantities to derived quantities to any other information required to restart a simulation from a specified point in time without any information being lost in sharing or analysis of results.

The ability of NIMROD’s MHD model to simulate the evolution of electrons as well as one or more species of ions at the same time will, in the output files, be reflected in the existence of distinct variables corresponding to each species in the plasma. For example, there could be distinct density and temperature variables corresponding to each species, as well as distinct velocity/momentum variables in the case of multiple ion species fluids. However, it’s crucial that a standardized output file format indicates which species each variable belongs to, as well as species-specific attributes such as charge/mass, where applicable, in order to allow proper attribution of each species’ contributions in the output file variables. Properly representing species information in a standardized way will become increasingly important as impurity ions are simulated in the plasma, as the standardized file format should be able to handle an arbitrary number of species without ambiguity.

NIMROD uses an unstructured two-dimensional finite element mesh in the poloidal plane and the Fourier method in the toroidal direction. The radial location can be in the form of the normalized poloidal flux, as it is aligned with the flux surfaces in the plasma. This will be helpful in the data interpretation. Finally, the domain of the plasma in the dumps will be extended beyond the LCFS into the SOL region. This means the dumps will contain data from the SOL region as well. Therefore, the standard form of the dumps from the NIMROD code should be such that it can be easily understood in space. This means the standard form of the dumps from the code should be able to give the location of the LCFS so that the data from the SOL can be distinguished from the core plasma data. The need to standardize the output of the NIMROD program stems from the following applications that have particular demands on the results:

\begin{itemize}
\item \textit{V\&V (Verification \& Validation):} For V\&V, consistency and completeness of metadata are paramount. A standardized output should include information about the run (e.g., equilibrium profiles used, key physical parameters, code version, units) in a well-defined way, so that another researcher can understand the conditions under which the data were produced. This makes it easier to reproduce results or compare simulation data with experimental measurements. Standardization also enables reproducible data extraction -- for instance, a common set of post-processing scripts or diagnostics can be applied to any NIMROD output if the format and naming are uniform. This helps ensure that verification checks and validation against experiments use the exact same definitions of quantities across different cases. In short, for V\&V purposes, the output files need to be self-describing and consistent from run to run, which a standardized format would enforce.
\item \textit{Coupling:} In integrated modeling or when coupling NIMROD to other codes, the outputs must be readily usable as inputs for the next tool. This requires a clear separation of the equilibrium (background) state and perturbations in the stored data, especially for linear simulations or evolving equilibrium cases. A standardized output could, for example, store the axisymmetric equilibrium fields separately from the 3D perturbation fields, allowing another code to import the equilibrium as an initial condition or to apply its own physics on top of it. Additionally, coupling benefits from using portable, non code-specific file formats (such as standardized HDF5 or NetCDF schemas) so that data exchange does not require custom converters. By adhering to common conventions for units, coordinates, and variable names, the NIMROD results can be more easily interpreted by different analysis codes or frameworks. In summary, a standard output format makes NIMROD data more plug-and-play for multi-code workflows, reducing the effort needed to translate or reinterpret the simulation results.
\item \textit{AI/ML:} For data-intensive tasks like machine learning, one might accumulate a large database of NIMROD simulation outputs (spanning many discharges or parameter scans). A uniform data schema is crucial in this context: every dataset should present the same structure so that \rev{an ingestion pipeline can automatically load the fields and profiles} without manual reformatting. Standard naming and organization of variables (for example, always using the same array names for density, temperature, etc., with the same dimensional layout) means that an ML algorithm can treat all cases uniformly. Moreover, standardization can facilitate feature extraction -- one could define a set of derived features (e.g., pedestal gradients, mode growth rates, energy confinement metrics) that are computed in a consistent manner for each snapshot and perhaps stored alongside the raw fields. This would enable efficient training of AI/ML models, since the inputs are well-defined and consistent. In essence, by making the output self-consistent and structured, we enable database-scale analyses and the application of modern data-mining techniques to NIMROD simulation data.
\end{itemize}

\begin{figure}[t]
\centering
    \includegraphics[width=1\linewidth]{combineNimrod.png}
\caption{\rev{Mapping of the NIMROD conversion workflow to IMAS, redrawn with separate static, nonlinear, linear, and metadata lanes to make the data paths unambiguous. Line colors follow the corresponding workflow boxes. In the linear case, \texttt{dump2imas} maps the final dump and \texttt{gamma2imas} stores growth rates and optional frequencies; in the nonlinear case, \texttt{dump2imas} is applied to the dump sequence.Color codes are used to differentiate the main lines of data communication: purple represents input data and configurations, blue is used for NIMROD pre-processing as well as simulation state communication, red is used for linear growth-rate and frequency communication, orange color is used for physical data going into IMAS IDS, and green is used for metadata and provenance. }}
\label{fig:mapping}
\end{figure}

\rev{The implementation described in this paper is intended for IMAS DD version 4.1 and beyond. This focus on a specific release and beyond follows two interrelated aspects: (i) non-backward-compatible IDS changes associated with the DD version 4 series~\cite{IMASDDv411}, and (ii) the DD v4 baseline coordinate convention COCOS=17, replacing the COCOS=11 convention used by DD v3~\cite{Sauter2013,IMASDDv411}.} While some of the ideas and techniques have broader applicability, the described implementation of the converter, the associated information population, and the coordinate/sign convention application should be considered centered on IMAS DD v4.1 and subsequent versions. 

The conversion workflow is a repeatable process that turns NIMROD inputs and internal NIMROD \texttt{dumpgll} HDF5 outputs into IMAS-compliant datasets as demonstrated in Fig.~\ref{fig:mapping}. Each conversion includes reading the experimental inputs and dump files and saving equilibrium and perturbation fields, followed by checking array and coordinates. The saved fields are then normalized and changed into IMAS unit conventions, which are defined in corresponding IDSs. The experimental inputs that are used by the NIMROD pre-processing are saved using a Python conversion script, \texttt{input2imas.py}. In addition to saving the experimental inputs, all input files to the NIMROD pre-processing \texttt{fgnimeq} code are saved in \texttt{equilibrium.code.parameters} of the \texttt{equilibrium} IDS, and the NIMROD input \texttt{nimrod.in} in the similar field of a dedicated \texttt{mhd} IDS. All information at this stage is saved in IDSs with an occurrence equal to zero. After NIMROD preprocessing is completed, the initial NIMROD \texttt{dumpgll.00000.h5} file is generated. For consecutive time steps, \texttt{dumpgll} files incorporate the numbers that correspond to individual time steps. Depending on the nature of the simulations, the dump files are saved at a frequency from 50 to 10000 time steps. The initial \texttt{dumpgll} as well as all consecutive dump files are converted using the \texttt{dump2imas.py} Python script, which mainly uses IMAS-Python for IDS creation and database I/O and \texttt{h5py} direct write for optimized access for selected variables and fields. The information from all \texttt{dumpgll} is saved in IDSs with occurrences greater than zero. The command-line interface records the target IMAS database name along with the \texttt{pulse} and \texttt{run} identifiers and the chosen data dictionary versioning plan, making sure consistent database layouts across installations and IMAS releases.

The software stack mixes Python orchestration with IMAS-Python \texttt{put()} actions for standard IDS filling. For big unstructured-grid products, where leaf-by-leaf IDS can become a main cost, the workflow supports a chosen fast path that writes specific datasets straight into the made IMAS HDF5 files via \texttt{h5py}. Basic checking is done using light utilities to confirm structural wholeness and the expected order of stored leaves. The conversion script shows user-changeable controls that set the resolution of stored products, including binning and reconstruction settings.

To accommodate device- and workflow-dependent conventions in NIMROD dumps, the converter uses clear multiplicative scale factors for key physical sizes, including pressure, temperature, density, magnetic field, length, and speed. These factors are offered as command-line options, allowing controlled mapping to IMAS main units without hard-coded settings. For example, the temperature is kept in IMAS in eV; so, the pipeline treats electron and ion temperatures as eV on output and scales the temperatures when the native dump display differs.

\subsection{COCOS transformation in NIMROD-to-IMAS converter\label{sec:cocos}}
NIMROD represents 3D fields in cylindrical geometry using a 2D finite-element discretization
in the $R$--$Z$ plane together with a Fourier representation in the toroidal direction $\phi$
(e.g., spectral toroidal harmonics). This structure is consistent with a natural cylindrical ordering
denoted $(R,Z,\phi)$ in NIMROD. In contrast, IMAS adopts the ITER right-handed cylindrical convention $(R,\phi,Z)$. 
The COCOS framework provides a compact identifier for the sign and
handedness conventions used in toroidal systems~\cite{Sauter2013}.
In this work, the NIMROD dump data are treated as COCOS=2, while the IMAS output is enforced to be
\rev{COCOS=17, which is the convention specified by IMAS DD v4.1~\cite{IMASDDv411}.} Operationally, the COCOS mapping corresponds to reversing the
toroidal angle direction (equivalently, changing the sign of the cylindrical handedness parameter
$\sigma_{R\phi Z}$), while keeping the physical $R$ and $Z$ directions fixed. This is equivalent to changing 
\rev{the sign of the cylindrical handedness parameter} $\sigma_{R\phi Z}$, which is a sign factor that specifies whether the cylindrical basis
$(\hat{\mathbf e}_R,\hat{\mathbf e}_\phi,\hat{\mathbf e}_Z)$ is treated as right-handed or left-handed. It is defined through the cross product
$\hat{\mathbf e}_R \times \hat{\mathbf e}_\phi = \sigma_{R\phi Z}\,\hat{\mathbf e}_Z$,
where
\begin{equation}
\sigma_{R\phi Z} =
\begin{cases}
+1, & \text{right-handed $(R,\phi,Z)$ convention},\\
-1, & \text{left-handed convention (equivalently, a reversal of the toroidal angle direction)}.
\end{cases}
\end{equation}
Changing the sign of $\sigma_{R\phi Z}$ is equivalent to reversing the toroidal basis direction,
$\hat{\mathbf e}_\phi \rightarrow -\hat{\mathbf e}_\phi$ (i.e., $\phi \rightarrow -\phi$), while keeping $R$ and $Z$ fixed. As a result, toroidal components of vector quantities (e.g., $B_\phi$, $V_\phi$, $J_\phi$) change sign under the mapping.

In the generic COCOS formulation, the equilibrium magnetic field can be written in terms of
poloidal flux $\psi$ and toroidal-field function $F$; importantly, the cylindrical handedness enters
the expressions for the poloidal components~\cite{Sauter2013}:
\begin{align}
B_R \propto \frac{\sigma_{R\phi Z}\,\sigma_{Bp}}{(2\pi)^{\epsilon_{Bp}} R}\frac{\partial \psi}{\partial Z},\quad
B_Z \propto -\frac{\sigma_{R\phi Z}\,\sigma_{Bp}}{(2\pi)^{\epsilon_{Bp}} R}\frac{\partial \psi}{\partial R},\quad
B_\phi = \frac{F}{R}.
\label{eq:cocos_BR_BZ}
\end{align}
\rev{In Eq.~\eqref{eq:cocos_BR_BZ}, $F$ is the signed toroidal-field function defined by $F=RB_\phi$~\cite{Sauter2013}; therefore an additional factor $\sigma_{R\phi Z}$ is not inserted in the expression $B_\phi=F/R$. For the COCOS 2-to-17 mapping used here, the toroidal component reverses sign and the signed field function changes accordingly, $F^{(17)}=-F^{(2)}$.}
The $\epsilon_{Bp}$ parameter takes values 0 for COCOS$<$10 or 1 for COCOS$>$10 and defines if the poloidal flux is normalized by the factor $2\pi$ or not. Changing $\sigma_{R\phi Z}$ would flip the signs of $B_R$ and $B_Z$ unless the sign of $\psi$ or $\sigma_{Bp}$ is also flipped. Likewise, reversal of the toroidal basis vector implies a sign change of toroidal vector components, including $B_\phi$, $V_\phi$, and $J_\phi$. In the COCOS mapping used here, $\sigma_{R\phi Z}$ and $\sigma_{Bp}$ change such that the product $\sigma_{R\phi Z}\sigma_{Bp}$ is preserved; therefore $\psi$ have the same sign and differ by the factor of $2\pi$, consistent with $\psi^{(17)} =2\pi \psi^{(2)}$.

Let superscripts $(2)$ and $(17)$ denote quantities expressed in COCOS 2 and COCOS 17,
respectively. \rev{The converter applies the following mapping before writing any IMAS IDS values; the resulting data and coordinate-system metadata are stored as COCOS=17, rather than storing COCOS=2 values for downstream reinterpretation:}
\begin{align*}
\psi^{(17)}(R,Z) &= 2\pi \psi^{(2)}(R,Z),\\
B_R^{(17)} &= B_R^{(2)},\qquad B_Z^{(17)} = B_Z^{(2)},\qquad B_\phi^{(17)} = -B_\phi^{(2)},\\
V_\phi^{(17)} &= -V_\phi^{(2)},\qquad J_\phi^{(17)} = -J_\phi^{(2)}, \label{eq:cocos_psi_flip}\\
\phi^{(17)} &= -\phi^{(2)}.
\end{align*}
with the poloidal components of $\mathbf{B}$, $\mathbf{V}$, and $\mathbf{J}$ unchanged. This conversion brings the toroidal direction and handedness into agreement with the target IMAS convention with COCOS=17.

NIMROD stores non-axisymmetric perturbations as toroidal Fourier harmonics. For a scalar perturbation
expanded as
\begin{equation}
\tilde{f}_n(R,Z,\phi,t) = \Re\!\left[ \hat{f}_n(R,Z,t)\,e^{i n \phi} \right],
\end{equation}
the toroidal-angle reversal $\phi^{(17)}=-\phi^{(2)}$ implies
\begin{equation}
\hat{f}_n^{(17)} = \left(\hat{f}_n^{(2)}\right)^{*},
\label{eq:cocos_scalar_conj}
\end{equation}
i.e. a complex conjugation of the mode coefficient. In the common storage form
$\hat{f}_n=\Re(\hat{f}_n)+i\,\Im(\hat{f}_n)$, Eq.~\eqref{eq:cocos_scalar_conj} corresponds to
$\Im(\hat{f}_n)\rightarrow -\Im(\hat{f}_n)$ with $\Re(\hat{f}_n)$ unchanged. The full perturbation can be obtained by an explicit sum over $n$.

For a vector perturbation
\begin{equation}
\tilde{\mathbf{A}}(R,Z,\phi,t)=\Re\!\left[\hat{\mathbf{A}}_n(R,Z,t)\,e^{i n \phi}\right], \quad
\hat{\mathbf{A}}_n=(\hat{A}_R,\hat{A}_Z,\hat{A}_\phi),
\end{equation}
the transformation includes both the scalar conjugation and the toroidal basis reversal,
\begin{equation}
\hat{\mathbf{A}}_n^{(17)}=\left(\hat{A}_R^{(2)\,*},\,\hat{A}_Z^{(2)\,*},\,-\hat{A}_\phi^{(2)\,*}\right).
\label{eq:cocos_vector_rule}
\end{equation}
If the converter stores real and imaginary parts separately, Eq.~\eqref{eq:cocos_vector_rule} implies:
\begin{align*}
\Re(\hat{A}_{R,Z})^{(17)}&=\Re(\hat{A}_{R,Z})^{(2)}, &
\Im(\hat{A}_{R,Z})^{(17)}&=-\Im(\hat{A}_{R,Z})^{(2)},\\
\Re(\hat{A}_{\phi})^{(17)}&=-\Re(\hat{A}_{\phi})^{(2)}, &
\Im(\hat{A}_{\phi})^{(17)}&=\phantom{-}\Im(\hat{A}_{\phi})^{(2)}.
\end{align*}
The converter applies these rules to the exported magnetic-field, flow, and current perturbations,
ensuring that the resulting IMAS mode data are consistent with COCOS=17.

Finally, the converter writes the IMAS coordinate-system metadata to indicate COCOS=17 for the
produced IDSs, so downstream IMAS tooling interprets the signs consistently with the IMAS
conventions.

\subsection{Conversion of NIMROD Finite-Element Mesh to IMAS GGD Unstructured Format}

The data output from the NIMROD simulations is in the form of plasma fields defined on a high-order finite element mesh in the poloidal $(R, Z)$ plane and Fourier expansions in the toroidal direction. This format is well-suited for spectral MHD solvers. However, it poses a challenge for inclusion in the IMAS framework due to its reliance on an explicit description of three-dimensional data. The General Grid Description (GGD), which is a part of the IMAS framework, allows for structured and unstructured grids. In this work, we choose the unstructured grid format as the most robust and interoperable means of encoding NIMROD output within IMAS. 

The IMAS GGD unstructured mesh representation makes the mesh explicit by using node coordinates and, where applicable, element connectivity. In principle, this method should preserve the original finite element topology used by the NIMROD mesh. This is accomplished by using the original node distribution and reproducible linearized connectivity suitable for visualization and topology-aware access. The current workflow includes an optional resampling onto user-defined toroidal planes and/or regularized poloidal sampling, which is used for machine learning. This method forgoes preserving the original solver mesh for a uniform tensor representation. \rev{Here ``nodes'' means the full high-order Lagrange nodal set, not only the corner vertices of each finite element. NIMROD represents fields with high-order Lagrange-type finite elements~\cite{Sovinec2004}; after shared element boundaries are stitched, the exported poloidal set therefore contains the polynomial-degree-dependent nodal degrees of freedom, including multiple points within each coarse element. Exporting this nodal set is not a reduction to the coarse element-vertex mesh. The optional regular-grid resampling is the path that introduces interpolation and can reduce spatial fidelity.}

An alternative route that has been recognized for its potential use in recent machine learning-based applications is the representation of the data from the $(R, Z)$ plane, as obtained from the finite-element method employed by the NIMROD code, onto a regular grid structure defined by the coordinates $(r, \theta)$. This type of data representation has been recognized as being conducive to structured arrays often utilized by neural network data structures~\cite{Rahman2024}. This data representation route has the advantage of facilitating the data formatting process and allowing for greater compatibility with neural operator-based data analysis tools, such as convolutional and Fourier-based neural operator tools, which are typically designed with regular grid data structures. However, this data representation route requires the use of an interpolation step from the high-order finite-element mesh onto the regular grid structure defined by the coordinates $(r, \theta)$. This step can lead to the introduction of artificial features and smoothing effects, particularly near the separatrices and edges. This data representation route does not utilize the GGD grid structure and, therefore, suffers from the aforementioned limitations. On the other hand, our method utilizes the unstructured GGD grid and therefore maintains the exact node locations obtained from the finite-element method. This route also allows for the representation of non-regularly shaped domains. Although the GGD data structure has a greater storage requirement and introduces greater complexity into the machine learning pipeline due to the unstructured nature and lack of regular indexing, this route has the advantage of providing a geometrically faithful representation and greater portability.

Although the NIMROD mesh is \emph{logically} structured (finite-element nodes in $(R,Z)$ combined with a discrete set of toroidal planes), the choice of an unstructured GGD representation changes how data are indexed and how locality is expressed. On a structured grid, field values are naturally stored as a multidimensional tensor,
\begin{equation}
Q_{i j k} \equiv Q(R_i, Z_j, \phi_k), \qquad i=1,\ldots,N_R,\; j=1,\ldots,N_Z,\; k=1,\ldots,N_\phi,
\end{equation}
so nearest-neighbor access and local operators follow immediately from index offsets, e.g.\ $(i\pm1,j,k)$ or $(i,j\pm1,k)$. Even when stored in a packed 1D array, the mapping between a linear index $p$ and $(i,j,k)$ is explicit and invertible once an ordering convention is known, for example
\begin{equation}
p \;=\; i + N_R\bigl(j + N_Z k\bigr), \qquad Q_p \equiv Q_{i j k}.
\label{eq:structured_packing}
\end{equation}
In an unstructured GGD grid, by contrast, values are attached to node IDs,
\begin{equation}
Q_p \equiv Q(\mathbf{x}_p), \qquad \mathbf{x}_p=(R_p, Z_p, \phi_p), \qquad p=1,\ldots,N_{\text{nodes}},
\end{equation}
and the integer label $p$ does not imply geometric adjacency or a tensor-product ordering. Locality must therefore be recovered from mesh topology (or geometric queries), typically through an explicit connectivity relation that lists, for each cell $c$, the vertex IDs
\begin{equation}
\mathcal{C}_c = (p_1,\ldots,p_m),
\end{equation}
which induces a neighbor set for a node $p$ such as
\begin{equation}
\mathcal{N}(p) \;=\; \{\, q \;|\; \exists\, c \text{ with } p\in\mathcal{C}_c \text{ and } q\in\mathcal{C}_c \,\}.
\label{eq:tneighbor}\end{equation}

The notation $\mathcal{N}(p)$ is the topological neighbor set that is often called the 1-ring neighborhood of node $p$
on an unstructured mesh. Eq.~(\ref{eq:tneighbor}) means that $q\in\mathcal{N}(p)$ if there exists at least one cell $c$ such that both $p$ and $q$ are vertices of that same cell. 

Trivial operations such as stencils, reshaping, and patch extraction on structured grids become topology-aware gather/scatter operations when working with unstructured grids. This is the way in which unstructured Generalized Grid Data (GGD) does not have a natural notion of $(i, j, k)$ indexing: the data is maximally explicit and geometry-faithful, yet tensor indexing and adjacency (spacing) information are not implicit and must be provided.

The main reason for converting the data from the output of the NIMROD simulations into a format that can be easily handled by the IMAS framework is the need for a data description that can be easily understood by machine learning algorithms. This is because the unstructured grid format of the GGD allows for an explicit description of node coordinates and data values. This removes any ambiguity in the description of data for machine learning algorithms and ensures that the data can be easily processed by machine learning algorithms.

The conversion begins by reconstructing the global $(R,Z)$ mesh from NIMROD's multiple poloidal blocks. This involves extracting nodal positions from the finite-element structure and assembling them into a unified 2D mesh. Next, the Fourier coefficients stored in the NIMROD dump file are read and used to reconstruct full 3D fields at a chosen number of toroidal angles $\phi_k = 2\pi k/N_\phi$. The number of toroidal planes is selected to resolve the highest retained toroidal mode, typically using $N_\phi \ge 2N + 1$ to prevent aliasing. For each toroidal plane, the real-space field values are recovered from the spectral representation using an inverse discrete Fourier transform:
\begin{equation}
F(R_i,Z_j,\phi_k) = F_0(R_i,Z_j) + 2 \sum_{n=1}^{N} \left[ \text{Re}(\hat{F}_n) \cos(n\phi_k) - \text{Im}(\hat{F}_n) \sin(n\phi_k) \right].\label{eq:fourier}
\end{equation}
This reconstruction yields a complete 3D dataset sampled on a logically structured mesh. \rev{Two representations are intentionally kept distinct. In \texttt{mhd\_linear}, the real and imaginary parts of each toroidal Fourier coefficient are stored on a two-dimensional poloidal representation, so the toroidal spectral content is retained without first sampling it onto GGD toroidal planes. In \texttt{mhd}, when a real-space 3D snapshot is requested, the stored harmonics are reconstructed at the selected $\phi_k$ planes and node-centered values are written on the 3D GGD. The current \texttt{mhd\_linear} export stores poloidal field samples rather than the native finite-element basis functions themselves; consequently, it is an analysis/interoperability representation rather than a bitwise NIMROD restart record, and the poloidal reconstruction fidelity is set by the stored 2D representation.}

To store this result in the IMAS structure, the GGD object is created and filled with unstructured node coordinates. A global index is assigned to each node, and the node locations in $(R, \phi, Z)$ are stored. Values are associated with node indices, consistent with standard unstructured-mesh semantics (as used in tools like Gmsh~\cite{Geuzaine2009}). The resulting mesh is a tensor product of the 2D FE node set and the toroidal angle. Although the original mesh is logically structured, the unstructured GGD format has the advantage of explicitly treating each node, which can simplify access and improve compatibility with other tools. Values are then assigned to the field values at each node based on the indices, and, if desired, the cell connectivities can be included to define the resulting volumetric elements. The options for the cell connectivity are described in the next subsection.

This process is accomplished via a Python utility script, \texttt{dump2imas.py}, which reads the NIMROD dump file and writes a fully compliant IMAS file using the HDF5 file format. Although the unstructured nature of the GGD representation results in increased verbosity and elevated I/O demands compared to the structured representation, the need to infer the mesh shape is removed, and the self-describing nature of the mesh is preserved, making the representation platform-independent. To alleviate the performance constraints, the node coordinates and connectivity array are precomputed using vectorized NumPy operations~\cite{Harris2020}, thus avoiding the need to loop at the Python level. This vectorization significantly accelerates the grid construction process by at least a factor of 2000.

In IMAS, the topological description of an unstructured mesh is kept in \texttt{grid\_ggd}, while field arrays kept in \texttt{ggd} refer to that topology by index. Specifically, \texttt{grid\_ggd} has one or more \texttt{grid\_subset}s, each representing a group of objects of a certain topological dimension, such as vertices or volume elements. Connectivity is stored under \texttt{grid\_subset/element/object/index}, where \texttt{element} says the element kind, such as triangle, square, or hexahedron, and \texttt{object/index} keeps, for each element instance, the list of vertex indices that define it. The vertex coordinates are stored separately in the \texttt{grid\_ggd/space} branch, which gives the $(R,Z,\phi)$ coordinates associated with the vertex subset. Each \texttt{ggd} then points to the right grid via \texttt{grid\_index} and shows if the values are on points or cells via \texttt{grid\_subset\_index}; this split lets the same mesh topology to be used again across many fields and helps downstream tools to avoid random triangulation by using the clear connectivity given by \texttt{grid\_ggd}.

\begin{figure}[htbp]
    \centering
    \includegraphics[height=6cm]{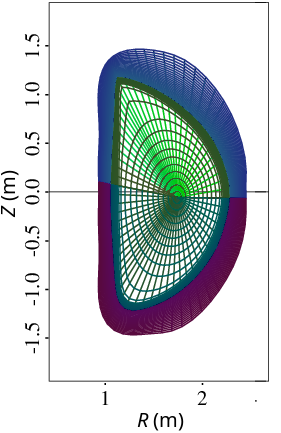} 
    \includegraphics[height=6cm]{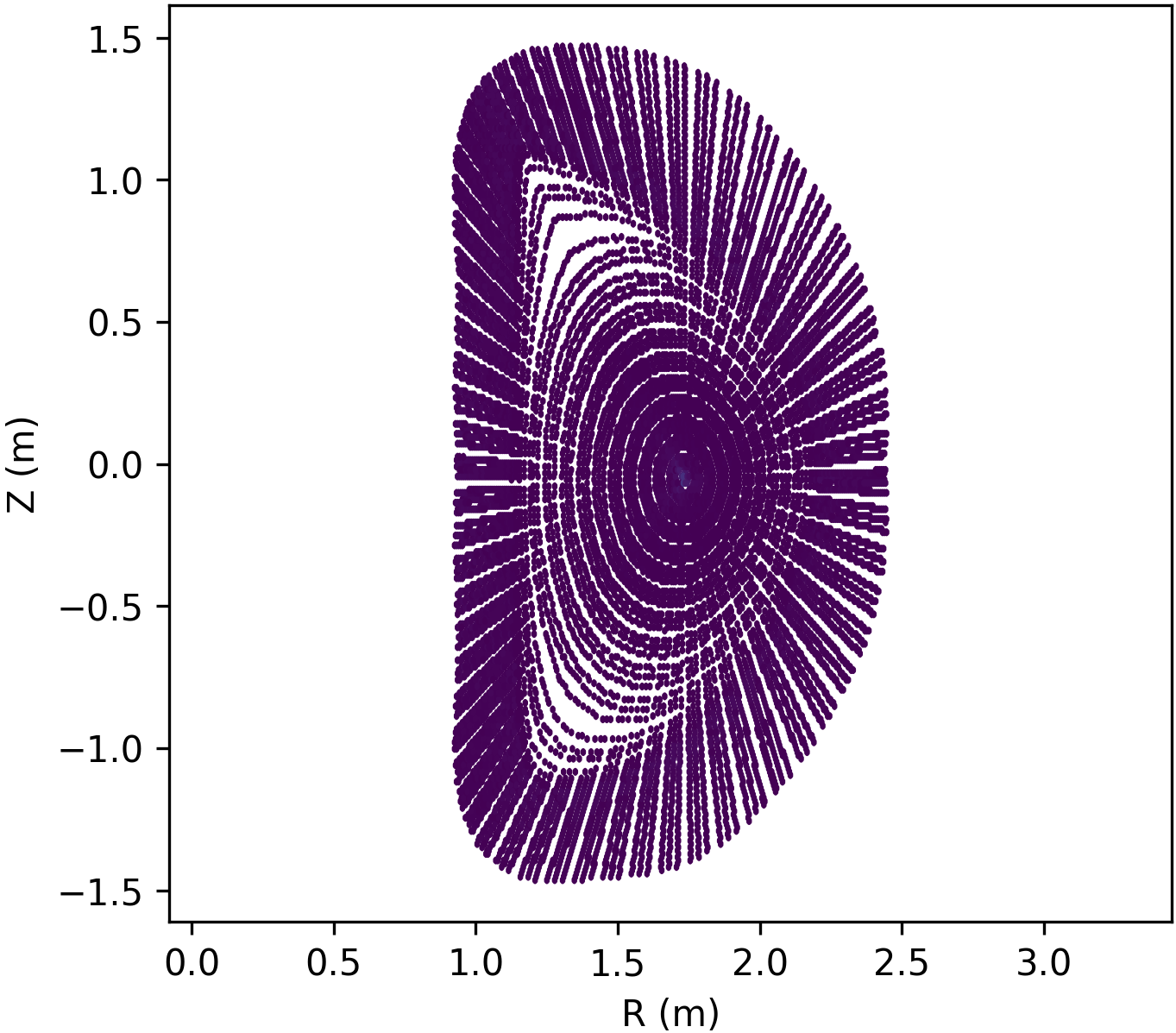}
    \caption{\rev{Comparison in the poloidal $(R,Z)$ plane. Left: NIMROD finite-element boundaries. Right: unstructured IMAS GGD representation showing the stitched high-order nodal set and triangular tessellation. Both panels are downsampled for visualization.}}
    \label{fig:mesh_comparison}
\end{figure}

\rev{Figure \ref{fig:mesh_comparison} shows an example of the transformation process in the poloidal plane. The left panel emphasizes the coarse finite-element boundaries, whereas the right panel is also a poloidal-plane view and shows the exported high-order nodal set after stitching and triangular tessellation. The additional points visible in the right panel are the polynomial-degree-dependent finite-element nodes within the coarse elements; they are not created by toroidal sampling. Each exported node is explicitly enumerated with its coordinates and associated values.}

By aligning the NIMROD data structure with IMAS's unstructured GGD standard, we create a format that is well-suited for modern data analysis and ML applications. This conversion ensures that field values are not only preserved with high fidelity but also encoded in a way that is resilient to version changes and backend differences in the IMAS framework.

\subsubsection{\rev{Poloidal nodal representation and unstructured connectivity options: \texttt{hex}, \texttt{fe\_tri}, \texttt{fe\_wedge}, and \texttt{fe\_pointcloud}}}
\label{sec:ggd_connectivity_options}

The IMAS GGD unstructured representation can include node positions and node-centered numbers. It can also include an explicit element connectivity that shows how nodes are grouped into cells. In the \texttt{dump2imas.py} process, this link is managed by the \texttt{ggd-connectivity} options implemented in the workflow. All these options are described below. They use the same node position rule: a tensor product of the poloidal $(R,Z)$ node set and a set of toroidal angles $\{\phi_k\}_{k=0}^{N_\phi-1}$. The differences between these choices are the volumetric cell topology used to represent the resulting three-dimensional domain. Some options are based on the native NIMROD mesh, while the others rely on the poloidal plane resampling. 

The first approach uses a uniform $(R,Z)$ resampling, followed by a hexagonal extrusion, to produce a logically consistent lattice within the poloidal plane (\texttt{hex})~\cite{ThompsonWarsiMastin1985}. This is achieved by resampling the computational domain onto a uniform $(R,Z)$ grid with user-specified bin number for each axis, effectively producing a structured lattice. This lattice is then extruded to produce 8-node hexahedral elements between the toroidal planes. The final connectivity is convenient for visualization, but this method does not preserve the natural packing of the finite element mesh from the NIMROD code. Therefore, this approach should be considered a utility approach rather than a true topology-preserving encoding of the solver's discretization.

The second option (\texttt{fe\_tri}) uses triangular tessellation~\cite{FreyGeorge2000,GeorgeBorouchaki1998,Shewchuk2002}. It preserves the native NIMROD finite-element nodal distribution in the poloidal plane by using the stitched global node set produced by assembling NIMROD blocks into a unified $(R,Z)$ mesh. A conforming triangular tessellation is then constructed in the poloidal plane. In the common case where the stitched lattice is logically quadrilateral, each logical quad cell is split into two 3-node triangles by a consistent diagonal. The triangular connectivity preserves the nonuniform resolution of the original discretization, including packing, avoids additional interpolation in the poloidal plane, and provides an explicit surface mesh suitable for consistent two-dimensional slicing and for generating volumetric elements by extrusion. Triangulated surface meshes are a standard and robust choice for representing unstructured geometries in scientific computing and visualization pipelines.  The connectivity for the toroidal triangular surface mesh that is based upon the \texttt{fe\_tri} representation includes both the triangles defined on each of the toroidal planes as well as the triangles created from edges connecting the toroidal planes of adjacent geometry locations. Therefore, the resulting triangular surface mesh is toroidally connected, but they do not represent a volumetric discretization. 

The third option extends \texttt{fe\_tri} to a volumetric unstructured mesh through the extrusion of the two-dimensional triangles between the adjacent toroidal planes. This approach makes use of the triangular-prism (wedge) extrusion, which is compatible with the (R,Z) × $\phi$ discretization (\texttt{fe\_wedge})~\cite{FreyGeorge2000,Hughes2000}. Each wedge (triangular-prism) cell has six nodes, corresponding to three nodes in one plane $\phi_k$ and the corresponding nodes in the adjacent plane $\phi_{k+1}$. \rev{The last of these planes is then connected back to the first with periodic closure along the toroidal direction}, resulting in a topologically closed 3D mesh. This approach is compatible with the product structure of the NIMROD outputs, where high-order finite elements are used in the $(R,Z)$ directions, and a discrete representation of a Fourier series is used in the $\phi$ direction, without altering the distribution of the poloidal nodes. This approach also avoids the necessity of a uniform poloidal lattice, which is necessary for the hex approach, and hence better preserves the regions that are refined in the solver’s mesh. This approach is also compatible with the conventional approach for the unstructured mesh, where prismatic elements are used for the extrusion of the triangular surface meshes. The \texttt{fe\_wedge} option is recommended for 3D volumetric visualization and volumetric post-processing, whereas \texttt{fe\_tri} is intended for connected surface representation and slicing.

Besides the three explicit connectivity options \texttt{hex}, \texttt{fe\_tri}, \texttt{fe\_wedge}, the NIMROD-to-IMAS converter offers a \texttt{fe\_pointcloud} option to export only the vertex subset of the unstructured GGD grid. In the \texttt{fe\_pointcloud} case, the stored node coordinates are the tensor product of the native NIMROD finite element nodal set in the poloidal plane $(R,Z)$ and a set of user-specified toroidal angles $\{\phi_k\}$. The volumetric element connectivity is not stored. Instead, field quantities are stored as node-centered values that reference the vertex subset via \texttt{grid\_index} and \texttt{grid\_subset\_index}. This representation retains the solver-driven packing strategy for the poloidal node set, including local refinement in regions of interest without the large storage and I/O penalty of the 3D cell connectivities. The \texttt{fe\_pointcloud} option is therefore suitable for workflows based on coordinates and values, such as the geometric neighborhood queries, feature detection, ML-based workflows, as well as a safe export option when connectivity-based extrusions would otherwise lead to excessive memory use or impractical IMAS file sizes. In cases where a tessellation is needed for a subsequent tool or algorithm, a 2D or 3D mesh can be constructed from the \texttt{fe\_pointcloud} later using a standard mesh generator without changing the physical values stored in the IMAS files.

\subsubsection{\rev{Toroidal indexing, periodicity, and node ordering}}
\begin{figure}[htbp]
\centering
\begin{tikzpicture}[scale=1.6, line cap=round, line join=round]
   \tikzset{
    pnode/.style={circle,fill=black,inner sep=1.1pt},
    lab/.style={font=\scriptsize},
    faceA/.style={draw=black, line width=0.35pt},
    faceB/.style={draw=black, dashed, line width=0.35pt},
    conn/.style={draw=black, line width=0.35pt},
    ghost/.style={draw=black!35, dashed, line width=0.3pt},
  }

  \def\dx{0.55}
  \def\dy{0.45}

  \begin{scope}[shift={(0,0)}]
    \draw[faceA] (0,0) rectangle (1.8,1.1);
    \node[pnode,label=below left:{0}]  at (0,0)   {};
    \node[pnode,label=below right:{1}] at (1.8,0) {};
    \node[pnode,label=below right:{2}] at (1.8,1.1) {};
    \node[pnode,label=below left:{3}]  at (0,1.1) {};
    \node[lab] at (0.9,-0.3) {$\phi_k$};

    \begin{scope}[shift={(\dx,\dy)}]
      \draw[faceB] (0,0) rectangle (1.8,1.1);
      \node[pnode,label=above left:{4}]  at (0.0,0.)   {};
      \node[pnode,label=above right:{5}] at (1.8,0) {};
      \node[pnode,label=above right:{6}] at (1.8,1.1) {};
      \node[pnode,label=above left:{7}]  at (0,1.1) {};
      \node[lab] at (1.0,1.3) {$\phi_{k+1}$};
    \end{scope}

    \draw[conn] (0,0) -- ++(\dx,\dy);
    \draw[conn] (1.8,0) -- ++(\dx,\dy);
    \draw[conn] (1.8,1.1) -- ++(\dx,\dy);
    \draw[conn] (0,1.1) -- ++(\dx,\dy);

    \draw[->,black!60] (2.06,0.05) -- (2.5,0.40) node[lab,anchor=west,black!60] {$+\phi$};

    \node[lab,anchor=north] at (1.1,-0.85) {\textbf{(a)} \texttt{hex} (8-node, 3D)};
  \end{scope}

  \begin{scope}[shift={(3.1,0)}]
    \draw[ghost] (0,0) rectangle (1.8,1.1);
    \node[lab] at (0.9,-0.3) {$\phi_k$};

    \begin{scope}[shift={(\dx,\dy)}]
      \draw[ghost] (0,0) rectangle (1.8,1.1);
      \node[lab] at (1.0,1.3) {$\phi_{k+1}$};
    \end{scope}
    \draw[->, black!60] (2.15,0.10) -- (2.55,0.43) node[lab,anchor=west,black!60] {$+\phi$};

    \coordinate (A) at (0.15,0.15);
    \coordinate (B) at (1.65,0.20);
    \coordinate (C) at (0.55,0.95);
    \draw[faceA] (A)--(B)--(C)--cycle;

    \node[pnode,label=below left:{0}]  at (A) {};
    \node[pnode,label=below right:{1}] at (B) {};
    \node[pnode,label=above:{2}]       at (C) {};

    \node[lab,anchor=north] at (0.9,-0.85) {\textbf{(b)} \texttt{fe\_tri} (3-node, 2D on one $\phi$ plane)};
  \end{scope}

  \begin{scope}[shift={(6.2,0)}]
    % bottom triangle (phi_k)
    \coordinate (A) at (0,0);
    \coordinate (B) at (1.8,0);
    \coordinate (C) at (0.9,1.25);
    \draw[faceA] (A)--(B)--(C)--cycle;
    \node[pnode,label=below right:{0}] at (A) {};
    \node[pnode,label=below right:{1}] at (B) {};
    \node[pnode,label=above:{2}] at (C) {};
    \node[lab] at (0.9,-0.3) {$\phi_k$};

    \begin{scope}[shift={(\dx,\dy)}]
      \coordinate (Ap) at (0,0);
      \coordinate (Bp) at (1.8,0);
      \coordinate (Cp) at (0.9,1.25);
      \draw[faceB] (Ap)--(Bp)--(Cp)--cycle;
      \node[pnode,label=below right:{3}] at (Ap) {};
      \node[pnode,label=below right:{4}] at (Bp) {};
      \node[pnode,label=above:{5}] at (Cp) {};
      \node[lab] at (0.9,0.7) {$\phi_{k+1}$};
    \end{scope}

    \draw[conn] (A) -- ++(\dx,\dy);
    \draw[conn] (B) -- ++(\dx,\dy);
    \draw[conn] (C) -- ++(\dx,\dy);

    \draw[->,black!60] (2.06,0.05) -- (2.55,0.45) node[lab,anchor=west,black!60] {toroidal $+\phi$};

    \node[lab,anchor=north] at (1.0,-0.85) {\textbf{(c)} \texttt{fe\_wedge} (6-node, 3D)};
  \end{scope}
\end{tikzpicture}
\caption{Connectivity conventions for optional unstructured cell topologies: (a) \texttt{hex} extrudes a uniform $(R,Z)$ lattice into 8-node bricks; (b) \texttt{fe\_tri} preserves the native FE node distribution and triangulates each logical quad cell; (c) \texttt{fe\_wedge} extrudes each triangle between adjacent toroidal planes to form a 6-node wedge cell. The node order is fixed to ensure reproducibility of cell orientation.}
\label{fig:connectivity_schematics}
\end{figure}
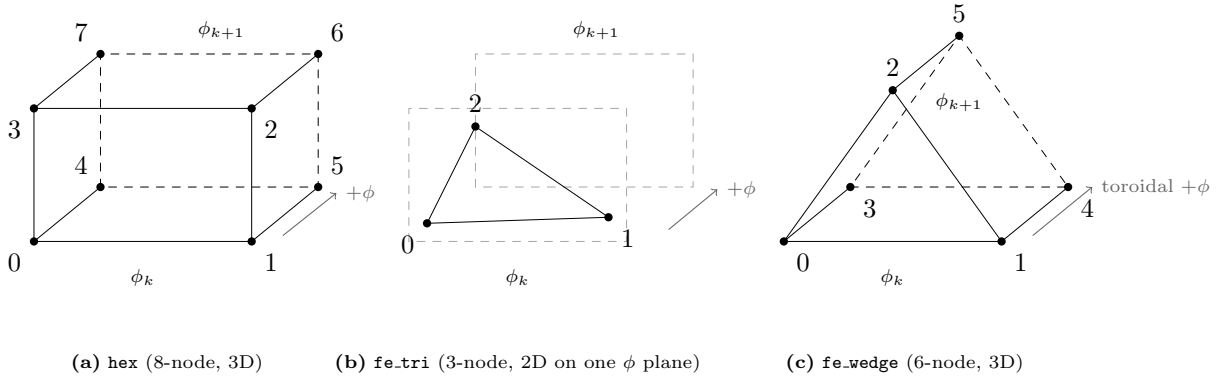

For all options, the node indexing is constructed as a product of poloidal node index $i\in[0,N_{2D}-1]$ and toroidal plane index $k\in[0,N_\phi-1]$, using a linearized convention
\begin{equation}
\mathrm{node}(i,k) = i + k\,N_{2D},
\end{equation}
with $\phi_k = 2\pi k/N_\phi$. In the volumetric options (\texttt{hex} and \texttt{fe\_wedge}), connectivity is formed between consecutive planes $(k,k+1)$. When periodic closure is enabled, the final plane is connected back to the first plane, $(N_\phi-1,0)$, producing a topologically periodic torus. For reproducibility and interoperability, the implementation uses a fixed local node ordering within each cell type (Fig.~\ref{fig:connectivity_schematics}); this is critical for downstream computation of cell volumes, normals, and consistent visualization of vector/tensor fields.

\subsection{Overall Conversion Workflow\label{sec:conversion}}
This data conversion process is carried out in two steps, as illustrated in Fig.~\ref{fig:mapping}. In the first, the static inputs and equilibrium fields are read in and mapped to the appropriate IMAS IDS structures with occurrence 0. In the second, the NIMROD simulation outputs (or ``dump'' files) are read in and appended to the appropriate IDS structures as additional time slices. In the actual process, the preprocessing step reads in the NIMROD equilibrium, profiles, and limiter geometry, which is written to the \texttt{equilibrium}, \texttt{core\_profiles}, and \texttt{wall} IDS structures, respectively, at time 0 (occurrence 0). The following time-stepped NIMROD outputs, which can include linear eigenmode solutions or fully nonlinear time-stepping solutions, are written to the \texttt{mhd\_linear} or \texttt{mhd} IDS structures (and to the \texttt{core\_profiles}, \texttt{core\_sources}, and others IDS structures, depending on whether time-stepping sources were included in the simulation) with the appropriate occurrence index greater than 0. \rev{Mesh-resolved real-space fields in \texttt{mhd} (and GGD-resolved edge quantities where used) are written with the IMAS General Grid Description as described in Sec.~\ref{sec:ggd_connectivity_options}. Mode-resolved Fourier coefficients in \texttt{mhd\_linear} are instead retained as two-dimensional harmonic arrays, as described below, so the spectral and real-space representations are not conflated.} 

While the use of the IMAS IDS label \texttt{mhd\_linear} might suggest a linearly related data structure, it is being utilized here as the natural home for the toroidal Fourier harmonics, which are a fundamental part of the NIMROD data representation. NIMROD contains the perturbations, and for nonlinear calculations, the full field content beyond ($n=0$), as a set of complex Fourier harmonics on a poloidal mesh, from which the full 3D field can be reconstructed by summation over $n$. \rev{Storing the harmonics in the \texttt{mhd\_linear} data structure maintains the toroidal spectral information and allows the 3D field to be reconstructed by recombination with the equilibrium data; the toroidal Fourier content is retained exactly, while poloidal fidelity is determined by the stored 2D representation.} Additionally, the reconstructed full 3D field data is stored on a 3D grid and written to the \texttt{mhd} IDS, providing a self-contained data structure for use by downstream codes that require direct access to the 3D field data.

Unit conversion and normalization follow the standard conventions appropriate to the physical quantities being represented, consistent with the IMAS standard. In the workflow, the equilibrium fields and static inputs, which are treated as constant in time, are established in the initial time slice, with the NIMROD dump files contributing to the dynamic IDS structures in the appropriate time slices. In linear NIMROD simulations, the initial and final time steps are generally written to the IDS structures, whereas in the course of a nonlinear simulation, there can be numerous time slices written to the IDS structures. This ensures that both linear and nonlinear NIMROD simulation outputs can be handled in the conversion process. 

\subsubsection{Equilibrium Mapping}
The \texttt{equilibrium} IDS represents a snapshot of the axisymmetric magnetic equilibrium and associated fields on a two-dimensional poloidal plane. In the present conversion, the equilibrium IDS is filled with information from NIMROD’s flux surface geometry and base fields. In other words, a snapshot of the poloidal flux function \(\Psi(R,Z)\)  and associated magnetic fields \((B_R, B_Z, B_{\varphi})\) on the IMAS grid in the \((R,Z)\) cross-section is recorded. Additionally, information on the last-closed-flux-surface (LCFS or separatrix) and magnetic axis positions is recorded. If a user-defined pressure profile is present in the input data, then associated profiles for pressure and corresponding current density is recorded in the equilibrium IDS as two-dimensional fields on a \((R,Z)\) grid. A snapshot of magnetic geometry and static profiles within a single axisymmetric configuration is represented. \rev{The \texttt{equilibrium} IDS assumes axisymmetric conditions and, therefore, three-dimensional information associated with toroidal perturbations is excluded from that IDS. This is a limitation of the axisymmetric scope of \texttt{equilibrium} for representing a non-axisymmetric state, but it is not a loss from the converted dataset because the perturbations are retained in \texttt{mhd\_linear} and, when requested, reconstructed in \texttt{mhd}.} 

\subsubsection{Profile Mapping: \texttt{core\_profiles} and \texttt{edge\_profiles}\label{sec:profiles}}
The \texttt{core\_profiles} IDS includes one-dimensional radial profiles of the plasma quantities, given by functions of the flux coordinate, typically the normalised toroidal flux. The input profiles for the NIMROD simulation, such as the density, temperature, and rotation, are mapped to this data structure. Each profile is given by an array for the various quantities, such as the electron density $n_e$, the ion densities $n_i$, the electron/ion temperature $T_e$/$T_i$, the toroidal/poloidal rotation, the $E\times B$ drift, etc. The data is given with respect to the normalised flux coordinate $\rho_{\rm tor}$, which ranges from 0 at the magnetic axis to 1 at the separatrix. The data is interpolated to a set of uniformly spaced or otherwise chosen radial coordinates, typically the NIMROD node points in the flux space. The units of the data are converted to the IMAS units, for example, the densities to $m^{-3}$, the temperatures to $eV$. \rev{If the simulation domain includes the scrape-off layer, the \texttt{core\_profiles} representation extends to the LCFS, while profiles beyond the LCFS are represented with \texttt{edge\_profiles}; the distinction is summarized in Table~\ref{tab:core_vs_edge_profiles}.} Since the toroidal flux is not defined in the SOL region, the profiles are stored as functions of normalized poloidal flux in the \texttt{edge\_profiles} IDS. In the case of a linear simulation, the core and edge profiles data are static. In the nonlinear simulation, the axisymmetric profiles ($n = 0$) can, in principle, be updated for all time slices. \rev{The \texttt{core\_profiles} and \texttt{edge\_profiles} IDSs contain one-dimensional profile data.} The \texttt{core\_profiles} for nonlinear simulations are post-processed by \texttt{dump2imas} workflow as flux-surface averaged one-dimensional profiles, functions of normalized toroidal flux. The coordinates for the data are the one-dimensional flux coordinate only. The ability to have the same one-dimensional profiles be used by other tools, such as TRANSP, for verification and validation, is a significant advantage. 

The NIMROD dumps are designed for code restartability and the field recovery and do not have information about explicit flux surface identifiers (e.g., \(\psi\) and \(\rho\)) nor data on the location of the LCFS. However, for IMAS to interoperate between different codes, it is essential to identify the location of the magnetic axis, establish a reference LCFS, and derive a normalized flux coordinate that separates the confined plasma from that in the SOL region.

The reconstruction of \(\psi(R,Z)\) is achieved through the use of the magnetic field components to assist in reconstructing the corresponding quantities. \rev{For example, to determine $\psi$ at a given point in the plane, we take a line integral of the equations below along a common path on the stitched, logically rectangular $(R,Z)$ grid using the magnetic-field components:} 
\begin{equation} 
B_R = -\frac{1}{2\pi R} \frac{\partial{\psi}}{\partial{Z}}\qquad B_Z = \frac{1}{2\pi R} \frac{\partial{\psi}}{\partial{R}}. 
\label{eq:psi}\end{equation} 
Line integration of this equation yields reconstructed \(\psi(R,Z)\). Eq.~\ref{eq:psi} is derived from Eq.~\ref{eq:cocos_BR_BZ} for COCOS=17 with $\sigma_{R\phi Z}=+1$, $\sigma_{Bp}=-1$, and $\epsilon_{Bp}=1$.

If an optional file, \texttt{contours.h5}, is available, it is used to obtain the LCFS contour. 
The converter reads these $(R,Z)$ coordinates of LCFS and interpolates \(\psi(R,Z)\) from the dump grid to these points, and then computes the reference LCFS \(\psi_{\mathrm{LCFS}}\) as the average:
\[\psi_{\mathrm{LCFS}} \equiv \text{median} \{ \psi(R_{\mathrm{LCFS}}(s_i),Z_{\mathrm{LCFS}}(s_i)) \}_{i=1}^{N_c}.\]
This method is accurate because it uses an explicit boundary to define the LCFS instead of relying on an interpolated boundary based on the field data.

If \texttt{contours.h5} is not available, the converter will use the IMAS \texttt{equilibrium} IDS occurrence 0 from the original pre-processed data, and the available values for $\psi_{\mathrm{axis}}$ and $\psi_{\mathrm{LCFS}}$ can be reused. If the \texttt{equilibrium} IDS occurrence 0 is not available as well, the converter reconstructs $\psi(R,Z)$ using Eq.~(\ref{eq:psi}) and finds $\psi_{\mathrm{LCFS}}$ by interpolating the electron temperature values at LCFS found both in the input \texttt{peqdsk} file and in equilibrium fields of NIMROD dump files. Finally, a similar approach is used for electron density fields as a last resort. This can be viewed as a quantile-based determination of the edge values for $\psi$, evaluated relative to the axis position. 

The converters compute and store 1D profiles by reducing stitched 2D fields \(Q(R,Z)\) onto flux-labeled radial coordinates. Formally, the flux-surface average of a scalar \(Q(\mathbf{r})\) on the flux surface labeled by \(\psi\) can be defined as a thin-shell volume average,
\begin{equation}
\left\langle Q \right\rangle_{\psi}
\equiv
\lim_{\Delta\psi \to 0}
\frac{\displaystyle \int_{\Delta V(\psi)} Q(\mathbf{r})\, d^3\mathbf{r}}
     {\displaystyle \int_{\Delta V(\psi)} d^3\mathbf{r}},
\qquad
\Delta V(\psi) \equiv V(\psi+\Delta\psi)-V(\psi),
\end{equation}
where \(V(\psi)\) is the volume enclosed by the surface labeled by \(\psi\). 
In practice, the NIMROD-to-IMAS converters approximate this in a discrete setting by binning values over a stitched \((R,Z)\) grid according to the local poloidal-flux label, computing bin means over points whose \(\psi_N(R,Z)\) fall within the same bin. The normalized poloidal flux coordinate is defined by
$\psi_N \equiv \left({\psi - \psi_{\mathrm{axis}}}\right)/\left({\psi_{\mathrm{LCFS}}-\psi_{\mathrm{axis}}}\right)\,$
and is related to the IMAS normalized poloidal-flux coordinate $\rho_{\mathrm{pol,norm}}=\sqrt{\psi_N}$. The  normalization for $\psi_N$ remains monotonic regardless of whether the absolute poloidal flux $\psi$ increases or
decreases from axis to LCFS for COCOS families with opposite $\Psi_{\rm ref}$ directions.

The IMAS \texttt{edge\_profiles} IDS is intended to represent plasma profiles within the edge/SOL region in a manner that is consistent with the presence of open field lines and SOL physics. Unlike \texttt{core\_profiles}, which are usually one-dimensional profiles on a flux surface and are confined within a core region, \texttt{edge\_profiles} is naturally applicable to profiles that extend across the separatrix into the SOL and PF regions. In particular, within the NIMROD-to-IMAS mapping used within this study, \texttt{edge\_profiles} stores one-dimensional profiles as a function of a normalized poloidal flux coordinate, $\psi_{\rm pol,norm}$, defined such that $\psi_{\rm pol,norm} = 0$ at the magnetic axis and $\psi_{\rm pol,norm} = 1$ at the LCFS, with $\psi_{\rm pol,norm} < 1$ corresponding to pedestal and PF regions both inside and outside the separatrix. This approach naturally avoids any artificial truncation of SOL region profile data due to the lack of a single monotonic radial coordinate that is valid across closed and open field lines. 

Although this method is very stable and does not require any explicit surface reconstruction, it may become geometrically ambiguous for regions where a single flux value corresponds to multiple spatially disconnected domains. For example, in the vicinity of and outside of LCFS, a single value of $\psi_{\rm pol,norm}$ may correspond to both the outboard pedestal/SOL and the PF region. Therefore, the bins with $\psi_{\rm pol,norm}> 1$ may have contributions from both regions simultaneously. This should be taken into account when interpreting the one-dimensional quantities in \texttt{edge\_profiles}; if a strict distinction between SOL and PF regions is desired, the GGD-resolved fields in \texttt{edge\_profiles} may be used in conjunction with geometric masks such as divertor leg or midplane selection.

The main practical distinction in implementation concerns the location of the mesh-resolved data, with the GGD branches used to hold the unstructured grid data being present only in \texttt{edge\_profiles}. This maintains the compactness of the \texttt{core\_profiles}, which hold 1D flux surface averages appropriate for a confined plasma, while \texttt{edge\_profiles} provide both (i) 1D profiles across core, edge, and SOL as functions of $\psi_{\rm pol,norm}$ and (ii) GGD data for spatially resolved data on the NIMROD mesh, relevant to edge-centric analysis and coupling. \rev{Differences between data stored in \texttt{core\_profiles} and \texttt{edge\_profiles} are summarized in Table~\ref{tab:core_vs_edge_profiles}.}

\begin{table}[t]
\centering
\caption{Summary of \texttt{core\_profiles} vs \texttt{edge\_profiles} usage in the NIMROD$\rightarrow$IMAS mapping.}
\label{tab:core_vs_edge_profiles}
\begin{tabular}{p{0.22\textwidth} p{0.36\textwidth} p{0.36\textwidth}}
\hline
 & \textbf{\texttt{core\_profiles}} & \textbf{\texttt{edge\_profiles}} \\
\hline
Primary 1D radial coordinate &
$\rho_{\mathrm{tor,norm}}$, with
$\rho_{\mathrm{tor,norm}}=0$ at axis and $=1$ at LCFS &
$\psi_{\mathrm{pol,norm}}$, with
$\psi_{\mathrm{pol,norm}}=0$ at axis, $=1$ at LCFS, and $>1$ in SOL/PF \\
\hline
Physical domain represented &
Confined plasma (core/pedestal inside LCFS) &
Full domain including core, LCFS, SOL, and private-flux region;
profiles outside LCFS are preserved (not forced to zero) \\
\hline
Content type (typical) &
Flux-surface averaged 1D scalar profiles
($n$, $T$, $p$, rotations, etc.) &
(1) 1D profiles across core/edge/SOL in $\psi_{\mathrm{pol,norm}}$;
(2) optional mesh-resolved quantities via GGD \\
\hline
GGD in this work &
Not used (kept lightweight for interoperability) &
Included; primary container for unstructured-grid fields and derived edge quantities \\
\hline
Typical consumers &
Core transport / integrated modeling workflows; confined-region initialization &
Edge/SOL analysis tools, visualization, coupling to edge models; boundary-condition interpretation \\
\hline
\end{tabular}

\end{table}

\subsubsection{Perturbation Mapping: \texttt{mhd\_linear}}
NIMROD uses a Fourier representation to represent perturbed MHD quantities. In this workflow, this representation is preserved and used to write the perturbations to IMAS files by mapping the perturbations to the \texttt{mhd\_linear} IDS. Although the IDS suggests a particular analysis type, this representation is also appropriate for nonlinear NIMROD runs; hence, complex perturbations are preserved for both linear and nonlinear runs if mode-resolved data are available. For each harmonic, the complex perturbation is preserved as separate real and imaginary components, as is appropriate for the representation of mode amplitudes in NIMROD. Modes are represented as separate values in the \texttt{toroidal\_mode[]} arrays, with the appropriate toroidal mode number held in \texttt{toroidal\_mode[i]\
n\_phi}. For each mode, the IDS represents the complex perturbation as two separate arrays (real, imaginary) on a poloidal plane at a specified toroidal angle, allowing the full perturbation to be reconstructed following the Fourier convention introduced earlier in Eq.~(\ref{eq:cocos_scalar_conj}). If the complex amplitude of the $n$th toroidal harmonic is written as
$\hat{f}_n = \delta f_{\mathrm{re}}^{(n)} + i\,\delta f_{\mathrm{im}}^{(n)}$, and 
\begin{equation*}
\Re\!\left[\hat{f}_n e^{i n \phi}\right]=
\Re\!\left[\left(\delta f_{\mathrm{re}}^{(n)} + i\,\delta f_{\mathrm{im}}^{(n)}\right)\left(\cos n\phi + i \sin n\phi\right)\right]=\delta f_{\mathrm{re}}^{(n)} \cos n\phi-\delta f_{\mathrm{im}}^{(n)} \sin n\phi.
\end{equation*} 

Then, the total perturbation takes the form:
\begin{equation}
\delta f(R,Z,\phi) \;=\; \sum_{k} \left[\delta f^{(k)}_{\mathrm{re}}(R,Z)\cos(n_k \phi) \;-\;
\delta f^{(k)}_{\mathrm{im}}(R,Z)\sin(n_k \phi)\right],
\end{equation}
where $n_k$ denotes \texttt{n\_phi} for the $k$-th stored mode.  This form avoids phase ambiguity and allows 
direct comparison to the original NIMROD Fourier coefficients. 

The eigenfunctions, as well as the perturbed fields, are represented on a 2D grid in the poloidal plane. The \texttt{dump2imas} converter projects the fields onto a regular $(R, Z)$ grid to enable visualization or further use. This mapping is done uniformly with respect to both scalar and vector perturbations. \rev{The \texttt{mhd\_linear} IDS supports time-dependent storage through \texttt{time\_slice} entries. For each additional NIMROD slice, the converter appends and populates one new \texttt{time\_slice} AoS entry; it does not preallocate or repeatedly resize one monolithic array containing all time slices. Existing entries remain unchanged while the next entry is populated.}

In order to ensure data integrity, the converter also carries out some basic checks on the data being written out, including:
\begin{itemize}
        \item Mode List Consistency: The stored set of \texttt{n\_phi} values is checked to ensure compatibility between slices. When the mode content differs, as in the case where different nonlinear mode sets are present, this is noted in the metadata/log file, and a separate \texttt{toroidal\_mode} list is written out with each slice.
        \item Grid Consistency: Grid in NIMROD is fixed, so that it can be reused for consecutive time slices. Nevertheless, to keep it consistent with the IMAS specification, the grid is saved for each time slice.
        \item Metadata Consistency: Units, component ordering, and field naming are all done uniformly, allowing time traces or mode-by-mode analysis without any special handling.
\end{itemize}

In addition to the perturbation fields, \texttt{mhd\_linear} has a place to save the growth rates and frequencies for individual toroidal modes. However, in contrast to the perturbation fields, this information is relevant only for linear NIMROD simulations. For linear NIMROD simulations, as shown in the top panel of Fig.~\ref{fig:mapping}, the \texttt{gamma2imas} converter is used in parallel with \texttt{dump2imas} to store growth rates and frequencies. The information is according to the IMAS schema location \texttt{time\_slice(it)/toroidal\_mode(i)/growthrate}.   That means, one growth rate value will be stored for each recorded \texttt{n\_phi}. Due to the fact that the \texttt{mhd\_linear} IDS data are naturally divided by \texttt{time\_slice} and \texttt{toroidal\_mode} entries, as well as the perturbation field, the tool \texttt{gamma2imas} assigns all scalar values across all existing \texttt{time\_slice} entries, ensuring that eigenvalue metadata are always accessible, regardless of the current \texttt{time\_slice} that is being viewed.
  
The growth rate calculation uses the standard NIMROD time history data files, such as \texttt{energy.bin} or \texttt{logen.bin}, and includes basic convergence diagnostics, including windowed averages, scatter/standard deviation, and the number of points, similar to the existing \texttt{nimpy} tool. This allows the user to assess the quality of the growth rate and decide if a particular mode has reached an exponential growth regime.
Moreover, \texttt{gamma2imas} has the option to write a mode frequency when an auxiliary NIMROD binary history file (\texttt{nimhist}) is supplied. In the NIMROD code, \texttt{nimhist} contains a time series for the complex mode amplitude at specified NIMROD code diagnostic locations. From this complex time series, an instantaneous complex frequency \((\gamma + i \omega)\) may be estimated using successive points in the series.
When \texttt{nimhist} is supplied, the estimated \(\omega\) is written to \texttt{time\_slice(it)/toroidal\_mode(i)/frequency}, with units as defined in the IMAS data dictionary, corresponding to the \texttt{n\_phi} field. In our convention, the sign of the mode frequency is defined according to the toroidal-angle COCOS convention. Since the NIMROD-to-IMAS mapping uses $\phi^{(17)} = -\phi^{(2)}$, the complex toroidal harmonic coefficient is conjugated, $\hat{f}_n^{(17)} = (\hat{f}_n^{(2)})^*$. The same argument applies to the complex probe signal used in \texttt{nimhist}. If the local complex mode amplitude is written as $A(t) \propto \exp\!\left[(\gamma + i\omega)t\right]$, then under the same assumption one obtains
\begin{equation}
A^{(17)}(t) = (A^{(2)}(t))^* \propto \exp\!\left[(\gamma - i\omega)t\right].
\end{equation}
Therefore, the growth rate $\gamma$ is unchanged, while the mode frequency changes sign. 

The primary restriction with this process is that the \texttt{mhd\_linear} schema currently does not have a place to store a spatial coordinate or coordinates of \texttt{nimhist} probes. This information might be flux surface dependent, and multiple frequencies might be associated in NIMROD with different probes. However, a typical NIMROD run has one probe file that allows monitoring frequencies at the area of interest, and its location can be restored with NIMROD input parameters, stored in the code sections of \texttt{mhd} and \texttt{mhd\_linear} IDSs. Other limitations of the IMAS DD schema are discussed in the next section.

\subsection{Coverage gaps in the \texttt{mhd} and \texttt{mhd\_linear} IDSs}
\label{sec:coverage_gaps}

A single snapshot of the NIMROD dump contains more information that can be represented in a clean fashion by the current IMAS \texttt{mhd} and \texttt{mhd\_linear} IDS definitions. The NIMROD-to-IMAS converter focuses on providing access to useful and schema-clean subsets of the necessary data product sets, rather than aiming for restart-quality completeness. These main gaps in IMAS DD relevant to the NIMROD output can be divided into three categories. \rev{The gaps below are those exposed by the NIMROD mapping: some are specific to \texttt{mhd}/\texttt{mhd\_linear}, while others are broader extended-MHD schema needs, and they should not be interpreted as limitations of every IMAS IDS.}

First, the representation of extended-MHD dissipation and closure information is not uniformly supported by the IDS structures. In particular, the NIMROD model is often run with spatially varying resistivity, viscosity, hyper-viscosity, and possibly other diffusion-like coefficients, as well as model-specific closures. While it is possible to indirectly communicate some of the relevant information through input data provenance or to infer it through combinations of stored fields and model selection, the IDS structure does not provide a comprehensive and standardized set of specialized nodes to accommodate these profiles and closure parameters.

Second, the multi-species physics supported by the current \texttt{mhd\_linear} IDS schema is incomplete. In particular, the NIMROD model is capable of evolving multiple ion species with their respective densities and temperatures, as well as additional two-fluid variables that might include pressure for different species. In contrast, the IMAS \texttt{mhd\_linear} IDS schema is largely organized to accommodate single fluid variables, and it does not provide any means to store perturbations and equilibrium state variables for any number of ion species in a way that reserves the full NIMROD model content. \rev{Multi-species support already exists in other parts of the DD: \texttt{edge\_profiles} uses species-indexed ion arrays in both one-dimensional and GGD branches, and SOLPS-ITER-to-IMAS work uses \texttt{edge\_profiles}/GGD for edge-plasma data~\cite{Penko2020SOLPS,IMASDDv411}. The gap identified here is specifically the absence of an analogous species-indexed perturbation representation in \texttt{mhd\_linear}. Any future extension should therefore be coordinated with existing IMAS species conventions and the relevant code communities, rather than introducing a NIMROD-specific competing layout.} 

Third, optional diagnostic information that is not always available in the IDSs. An illustrative case is the mode frequency information derived from a location-specific complex amplitude history diagnostic, as discussed in the previous section. Similar issues arise for other run-dependent diagnostics and internal solver states that are useful for analysis, but are not necessarily needed for the analysis of the results.

The consequence is that, if NIMROD were to rely on IMAS as a native I/O and restartability, the current \texttt{mhd}/\texttt{mhd\_linear} schema would be insufficient to meet the requirement of restartability across the entire range of extended MHD models and options. At the same time, such information is not necessarily critical to the primary objective emphasized here: standardized, machine-readable datasets for validation, integrated modeling interfaces, and database-scale AI/ML applications. For example, the presence of transport-like coefficients such as effective thermal conductivity could be reconstructed using (i) the saved fields and profiles in the \texttt{mhd}/\texttt{mhd\_linear} dataset and (ii) the saved NIMROD input parameters in the IDS code section, even if the profiles of the coefficients themselves are not saved as dedicated IDS leaves.

Thus, the present workflow should be understood as an interoperable and analysis-oriented representation of NIMROD outputs, rather than as a universal replacement for native restart dumps.

\subsection{Workaround Strategy for Coverage Gaps\label{sec:gaps}}
\label{sec:workaround_strategy}

In order to fill this coverage gap without compromising schema cleanliness, the converter uses a conservative approach that focuses on two areas: (i) consistency in organization for analysis tools, and (ii) minimization of duplicated high-volume arrays.

The primary workaround for this issue of multi-species is the utilization of multiple IDS occurrences for storing species-specific scalar content. Occurrence ``0'' is reserved for the ``experimental'' dataset used in NIMROD preprocessing tools such as \texttt{fgnimeq}. Occurrence ``1'' of the \texttt{mhd\_linear} IDS holds information related to electrons. Remaining occurrences (greater than 1) are used for additional species information, such as ion densities and temperatures. \rev{However, vector perturbations represented only once in the current mapping---for example, magnetic-field and current perturbations, and the bulk-flow perturbation where applicable---are saved in a single occurrence to avoid duplicating large arrays.} This approach is explicitly defined in terms of species ordering, naming, and metadata tags in order to allow scripts to utilize this approach deterministically. Since the \texttt{mhd} IDS includes leafs for storing information related to individual ions, no additional occurrences are necessary.

This approach is intentionally pragmatic in nature. However, this approach also includes several trade-offs, which include limited discoverability through generic IMAS tools.  Users need to understand that species are organized in separate occurrences of \texttt{mhd\_linear}. However, incompatibility with other tools is a potential concern, which is addressed in this approach with good documentation. This is a concern that is easily managed in a controlled environment, but which provides further scope for future additions to the data dictionary, such as native multi-species support and standard dissipation/closure fields.

A second, implementation-level workaround is the mixed I/O approach used to make large-scale conversions practical for the HDF5 backend. The approach uses IMAS-Python for standard IDS creation and database I/O, while providing a fast path for large arrays by directly writing specific datasets to the IMAS HDF5 layout via h5py block operations. Light-weight structural checks are performed to ensure that the resulting files remain structurally consistent with the expected IDS layout. In the present implementation, IMAS-Python access for leaf-by-leaf population is at least an order of magnitude slower (observed slowdowns of over a factor of 2000) than direct h5py block writes for large arrays; therefore, h5py is used as a default approach for high-volume arrays such as connectivity arrays for unstructured GGD. 

\rev{We retain the IMAS-native writer for standard IDS population because it is the preferred path for long-term schema stewardship: it follows the active DD rather than hard-coding backend leaf paths. Table~\ref{tab:performance} compares this path with the direct \texttt{h5py} block-write optimization for the large unstructured-grid payload. For the point-cloud case, the two paths differ by only about 3\%. When explicit connectivity is included, the tested IMAS-native path is approximately 2200--3200 times slower than the direct block write, as summarized in the table. For the direct block write, the wedge case is about 1.5 times slower than the triangle case, consistent with its larger connectivity payload.}
\begin{table}[t]
  \centering
\caption{Wall-clock time (s) for writing unstructured GGD topology and a representative set of node-centered fields using the IMAS-native IMAS-Python path and the direct \texttt{h5py} block-write path to the IMAS HDF5 backend. The representative case described in Sec.~3.1 was converted for two time slices. The benchmark was performed on a Linux workstation equipped with an AMD Ryzen 9 3950X processor (16 physical cores and 32 hardware threads) and 62~GiB of usable memory. The benchmark files were stored on a local ext4 filesystem on a 6~TB Seagate ST6000DM003 hard disk. The execution environment has \texttt{OMP\_NUM\_THREADS=32}. The software environment used Python 3.12.3, \texttt{dump2imas} v0.3.3, IMAS-Python v2.1.0.post1, IMAS DD 4.1.1, NumPy 2.4.1, and h5py 3.15.1. The reported time is the elapsed wall-clock time for the complete \texttt{dump2imas.py} conversion of the two selected time slices, including GGD construction, IDS population, and HDF5 output. Absolute timings depend on hardware, filesystem/storage performance, software versions, and selected conversion options.}  \label{tab:performance}
  \begin{tabular}{lrr}
    \hline
    \textbf{GGD option} & \textbf{IMAS-native (s)} & \textbf{\texttt{h5py} block (s)} \\
    \hline
    \texttt{fe\_pointcloud} & 73.30 & 71.01 \\
    \texttt{fe\_tri}        & \rev{640,340.31} & 198.66 \\
    \texttt{fe\_wedge}      & \rev{663,059.75} & 300.53 \\
    \hline
  \end{tabular}
\end{table}

This behavior is consistent with the cost model for IMAS-Python population of array-of-structures (AoS)
leaves, where connectivity is realized through a large number of Python-level assignments into nested
IMAS containers. A simple estimate of the assignment count for topology construction is
\begin{equation}
N_{\mathrm{assign}} \approx 3N_{\mathrm{nodes}} + N_{\mathrm{verts}}N_{\mathrm{cells}},
\end{equation}
where $N_{\mathrm{nodes}}$ is the number of mesh vertices, $N_{\mathrm{cells}}$ is the number of mesh 
cells stored in the connectivity, and $N_{\mathrm{verts}}$ is the number of vertices per cell with 
$N_{\mathrm{verts}}=3$ for triangles, $6$ for wedges, and $8$ for hexahedra. For the
representative NIMROD mesh used in Sec.~\ref{sec:eho_case} of this paper with $N_r\times N_\theta = 72\times 128$ finite elements in the
poloidal plane with polynomial \rev{degree of 5 for the stitched nodal representation, the} nodal lattice has
\begin{align}
N_{r,\mathrm{nodes}} &= N_r\,p + 1 = 72\cdot 5 + 1 = 361, \\
N_{\theta,\mathrm{nodes}} &= N_\theta\,p + 1 = 128\cdot 5 + 1 = 641,
\end{align}
giving $ N_{2\mathrm{D}} = N_{r,\mathrm{nodes}}\,N_{\theta,\mathrm{nodes}} = 361\times 641 = 231,401$
nodes per toroidal plane. If each logical \rev{quadrilateral in the stitched nodal lattice is split into two triangles, the resulting two-dimensional topology contains $N_{\mathrm{cells}}^{2D}= 2(N_R-1)(N_Z-1)= 460800$ triangles per toroidal plane. The \texttt{fe\_tri} representation stores this triangular topology independently on each toroidal plane, whereas the \texttt{fe\_wedge} representation connects corresponding triangles on adjacent toroidal planes to form six-vertex wedge elements. The estimated numbers of topology assignments therefore scale as
\[
N_{\mathrm{assign}}^{(\mathrm{tri})}
    \approx 2.1\times10^6 N_\phi ,
\qquad
N_{\mathrm{assign}}^{(\mathrm{wedge})}
    \approx 3.5\times10^6 N_\phi .
\]
For the performance test, $N_\phi=4$ toroidal planes were used, giving approximately $8.3\times10^6$ and $13.8\times10^6$ assignments, respectively. Their ratio is approximately 1.67, consistent with the \texttt{fe\_wedge} direct-\texttt{h5py} write taking about 1.5 times longer than the \texttt{fe\_tri} write} in Table~\ref{tab:performance}. 

\rev{For the IMAS-native path, the wedge/triangle timing ratio is about 1.04, indicating that both variants are dominated by Python/IMAS object-management overhead rather than by the payload size.}

In situations where the grid topology is static over time, the object management overhead can be mitigated by writing the connectivity data once and reusing it for multiple time steps, thereby reducing the overall time by a factor proportional to the number of slices avoided. \rev{These measurements characterize the post-processing conversion of stored NIMROD dumps; they are not a comparison with the wall-clock cost of a single NIMROD time advance and do not change the simulation time step. The practical benefit is campaign-scale processing of hundreds or thousands of saved dumps and interactive turnaround for downstream analysis. Absolute timings, and potentially the observed ratios, depend on the machine, filesystem, Python/IMAS versions, and storage backend. The table therefore documents one tested platform/configuration rather than a hardware-independent performance ratio; broader cross-platform benchmarking is left for future work.}

\rev{The direct \texttt{h5py} path is not intended to define a non-IMAS data model: for the tested DD version it writes the same DD-defined HDF5 leaf paths and the result is checked by IMAS readback/structural validation. The important distinction is version stewardship. Because the low-level path bypasses the native writer, it does not automatically inherit a future DD leaf reorganization; the converter must be updated or the entry migrated/re-written for the new DD. By contrast, using the native IMAS writer avoids this backend-path dependency and is therefore preferable when performance is acceptable. We do not assume that a generic post-hoc \texttt{h5py} edit can automatically restructure arbitrary future DD changes. An official high-throughput bulk-write mechanism in the IMAS access layer would be the preferred long-term solution, combining native schema handling with block-I/O performance. The dataset records both DD and converter versions so that any version-specific fast-path layout can be detected and validated.}

\subsection{Provenance, FAIR metadata, and workflow recording~\label{sec:fair}}
\label{sec:provenance}

The reproducibility of the integrated modeling workflows is ensured with the following conditions being met for every entry in the IMAS: (i) a brief and human-readable dataset description; (ii) dataset metadata following the FAIR guidelines (e.g., dataset identifier, title, contacts, rights, etc.); and (iii) a machine-readable record of the processing steps that led to the creation of the IDSs stored. The above requirements are achieved with the NIMROD-to-IMAS converter via the use of the data structures summary, \texttt{dataset\_fair}, and workflow, as described in this paper (see Fig.~\ref{fig:mapping}).
  
The Python script \texttt{input2imas.py} produces a summary IDS that includes the following: machine name, pulse number, dataset description, and a provenance note contained in \texttt{ids\_properties.comment}. The above summary is intended to be light and DD4 compatible to provide a ``front page'' for the dataset that can be accessed without the need to open the associated physics IDSs.
  
Dataset metadata following the FAIR guidelines is stored in the \texttt{dataset\_fair} IDS. The metadata is derived from a user-provided YAML metadata file (e.g., nimrod.yaml). The IMAS factory function \texttt{build\_dataset\_fair\_ids()} constructs the \texttt{dataset\_fair} IDS and populates the dataset identifier, title, description, contact/provider, and licensing information from the YAML metadata file, where present. In addition, the function appends a note to the \texttt{ids\_properties.comment} section.
  
To the above metadata information, the converter appends a condensed record of the execution steps to the \texttt{dataset\_fair.ids\_properties.comment} section. For each step of the conversion process, the manifest checksum can be computed from the checksums of the input files and appended to the step description.
  
Each of these stages, \texttt{input2imas}, \texttt{dump2imas}, and \texttt{gamma2imas} creates a machine-readable log entry comprising the component's name, e.g., \texttt{nimrod2imas:dump2imas}, a UTC timestamp, a sanitized command-line input, and a list of input files, along with a hash for each file if enabled. Hash generation is controlled via the command-line interface, where it is enabled by default with the algorithm specified by the \texttt{checksum-algorithm} option. Hash generation is also controlled via metadata for automated execution.
  
As Python execution of the IMAS workflow IDS generation is restricted, the converter stores the \texttt{workflow} IDS, stored in the IMAS entry directory. Internally, this file uses a \texttt{time\_loop\&component[]} table to store information comprising the component's identity, i.e., its name, merged parameters, and optionally description, repository, and version. Executions are stored as XML and appended to \texttt{time\_loop\&component[]\&executions}. Hence, multiple calls to the same stage, e.g., multiple calls to \texttt{dump2imas} for different nonlinear time slices, are recorded without overwriting previous records. In this way, this workflow record provides the authoritative provenance for the entire conversion pipeline, and \texttt{dataset\_fair} provides a standard FAIR metadata container and a concise integrity report.
  
The provenance recording is independent of the underlying linear or nonlinear physics and records any executed stages in the conversion pipeline. In the case of a linear simulation, the log entry for the workflow might comprise \texttt{input2imas} and a single \  texttt{dump2imas} stage. In the case of a nonlinear simulation, multiple \texttt{dump2imas} stages might be recorded, corresponding to multiple time indices/occurrences, and a single \texttt{gamma2imas} stage for stability analysis product generation.

\section{Verification, validation, and performance}

In this section, the results of running the conversion pipeline on the example nonlinear NIMROD simulation run for DIII-D shot 163518 are provided. The goals are to describe the format of the resulting IMAS records, to verify the retention of the fundamental data concerning the equilibrium, profiles, and perturbations, and to provide information about validation tests that have been performed on the converted data. The submitted dataset should be considered an example converted record, not an exhaustive collection of runs.

\subsection{Example conversion of DIII-D discharge 163518}
\label{sec:eho_case}
  
In a tokamak that operates in enhanced confinement mode (H-mode) confinement, a steep edge pressure pedestal in the plasma edge forms, and it often leads to periods of repetitive magnetohydrodynamic instabilities known as edge localized modes (ELMs). ELMs cause rapid losses of plasma particles and energy into the scrape-off region and divertor plates. In large-scale devices, ELM-induced transient plasma fluxes are generally considered one of the major concerns for plasma-facing component longevity and steady-state reactor operation, which may cause rapid erosion and melting of divertor plates and first-wall surfaces. Therefore, one of the primary objectives of integrated edge pedestal physics is to develop and validate steady-state confinement scenarios that maintain high confinement properties without ELMs or with ELMs replaced by stationary edge plasma transport processes.
  
The quiescent H-mode (QH-mode) is one prominent example of ELM-free scenarios that maintain H-mode confinement properties without large ELM losses due to continuous edge plasma transport driven by saturated edge fluctuations with low toroidal mode numbers. These edge fluctuations often take the form of edge harmonic oscillations (EHOs), which provide a means for stationary control of pedestal gradients. The EHO simulations used in this work (Ref. ~\cite{Pankin2020}) provide a relevant and challenging test case for archiving nonlinear extended MHD data in IMAS due to well-resolved time-dependent behavior that requires adequate storage rates for quantitative comparison with experimental and synthetic diagnostic data.
  
Our previous NIMROD study reported results of our simulations using the NIMROD code for the DIII-D discharge 163518, with the experimental equilibrium at $t\approx 2350~ms$~\cite{Pankin2020}. In linear stability NIMROD analysis, the equilibrium was found to be stable. In order to access the nonlinear EHO regime, an additional instability drive was provided by reducing the pedestal width, creating equilibria with pressure gradients increased by factors of 1.4 and 2.0 compared to the experimental reconstruction.  The nonlinear dynamics produced edge-localized, rotating perturbations with low toroidal mode numbers, allowing for the detailed validation of the simulations against experimental fluctuation measurements.
  
A critical part of the validation process is the determination of the correspondence between the simulated results and the experimental data, as obtained by the beam emission spectroscopy (BES). Let $I(t)$ and $I_{\mathrm{ref}}(t)$ denote the time traces from two BES channels (or a channel and a chosen reference). The cross-correlation used to quantify coherent rotating structure can be written as
\begin{equation}
    J_{\mathrm{cr}}(\tau) \;=\; \sum_{n=1}^{N} I(t_n+\tau)\,I_{\mathrm{ref}}(t_n),
    \label{eq:bes_xcorr}
    \end{equation}
    and its Fourier transform gives the cross-power spectral density,
    \begin{equation}
    S(f) \;=\; \mathcal{F}\!\left\{J_{\mathrm{cr}}(\tau)\right\},
    \label{eq:bes_cpsd}
\end{equation} 
which provides a compact characterization of dominant frequencies and phase relationships across channels.

In the present study, the validated EHO dataset serves as a stress test for the robust and analysis-ready IMAS archival, as nonlinear EHO simulations demand frequent output to resolve mode rotation and intermittency. In the DIII-D 163518 EHO production simulations, each NIMROD output is of the order of 1.3 GB. While the total number of dump files from the NIMROD simulation includes order to $10^3$ files, only 213 files are selected for conversion to the IMAS dataset. These dump files were previously used for synthetic diagnostic analysis and are the most relevant to be used for future ML/AI analysis. For the conversion of dataset of 276~Gb, I/O efficiency and schema correctness remain very important.
  
In the NIMROD to IMAS conversion, the perturbations are stored in the \rev{\texttt{mhd\_linear}} IDS using complex Fourier harmonics in the toroidal angle, $\hat{F}_n(R_i,Z_j,t)$, as well as equilibrium components. The 3D field on a toroidal grid, $\phi_k$, can be reconstructed for visualization and comparison with diagnostic geometry by the formula given by Eq.~\ref{eq:fourier}. In addition, the entire geometric information is also preserved by storing the generalized grid description (GGD) as an unstructured mesh, including node coordinates and element connectivity. In the case of volumetric elements, this includes the vertex index lists for the standard element types, such as tetrahedra, hexahedra, wedges, and pyramids, and optionally, in the case of nodal data only, a ``point-cloud'' connectivity representation when only nodal quantities are needed.  Maintaining complete GGD connectivity is essential for reproducible interpolation, conservative integration, and interoperability with unstructured-mesh post-processing and synthetic diagnostics.

Current IMAS coverage for the conversion process can be found in Table~\ref{tab:coverage}. Explicit storage of dissipative profiles, which include the resistivity, conductivity, and viscosity, is one area where IMAS is currently lacking. The absence of explicit storage of these is because there is currently no single defined location in the existing IMAS data model for these extended MHD simulation parameters, so that they can remain interoperable across MHD simulation workflows. However, from a machine learning (ML) viewpoint, these parameters can be reconstructed from the original inputs combined with the stored density and temperature coefficient profiles available in the IMAS infrastructure using standard model assumptions such as Spitzer-type resistivity dependence, \rev{$\eta\propto Z_{\rm eff}\ln\Lambda/T_e^{3/2}$}, and in this case, the lack of an explicit storage of the dissipative profiles does not hinder the ability to create features consistently, as long as the provenance records contain the original input decks and other model options.

\begin{table}[bh!]
\centering
\caption{NIMROD-to-IMAS conversion coverage relevant to AI/ML dataset requirements.}
\label{tab:coverage}

\begin{tabular}{p{0.12\linewidth} p{0.38\linewidth} >{\raggedright\arraybackslash}p{0.16\linewidth} p{0.26\linewidth}}
\hline
Stage & Data persisted in IMAS & Primary IDSs & Notable gaps / notes \\
\hline
\texttt{input2imas} &
Equilibrium and kinetic profiles derived from GEQDSK/PEQDSK; wall/limiter outline; preservation of NIMROD-facing namelists as XML in \texttt{code.parameters}; compact dataset metadata (summary/FAIR metadata) &
\texttt{equilibrium}, \texttt{core\_profiles}, \texttt{wall}, \texttt{summary} &
Depends on available inputs; intended as the canonical record for baseline equilibrium/profiles and input provenance \\
\texttt{dump2imas} &
Time-dependent equilibrium/profile content from dump files; mode-resolved perturbations on stitched grids; optional GGD representation of field-level snapshots for \rev{high-fidelity mesh-native analysis} workflows &
\texttt{equilibrium}, \texttt{core\_profiles}, \texttt{edge\_profiles}, \texttt{mhd\_linear}; \texttt{mhd} (for nonlinear simulations only) &
Dense field payloads can be selectively stored; some quantities may be mirrored or reconstructed to satisfy schema constraints depending on DD version \\
\texttt{gamma2imas} &
Mode scalars derived from diagnostics (growth rate; optionally frequency) stored with the corresponding toroidal modes &
\texttt{mhd\_linear} &
Requires diagnostic time histories; stores reduced scalars rather than full field histories \\
All stages &
Workflow provenance per step (component identity/version/repository, command, optional checksums) and dataset FAIR metadata for auditability and reuse &
\texttt{workflow}, \texttt{dataset\_fair} &
Enables version-aware ML retraining and dataset splits based on metadata filters \\
\hline
Missing (intentional) &
Explicit dissipation profiles (e.g.\ resistivity, conductivity, viscosity) &
--- &
No IMAS location for MHD dissipation profiles \\
\hline
\end{tabular}
\end{table}

\iffalse
Table~\ref{tab:data-inventory} summarizes the main categories of dataset records available through Princeton Plasma Commons as it is described in the Data Availability Section at the end of this paper. 
\begin{table}[h!]
\centering
\caption{Summary of deposited data records in Princeton Data Commons (PDC).}
\label{tab:data-inventory}
\begin{tabular}{p{0.26\linewidth} p{0.50\linewidth} p{0.18\linewidth}}
\hline
\textbf{Record type} & \textbf{Description} & \textbf{Format} \\
\hline
IMAS entry archives &
Converted NIMROD simulation datasets stored as IMAS entries; include equilibrium, profiles, wall, and
MHD (linear or nonlinear) IDSs, plus metadata/provenance IDSs. &
Compressed archive of IMAS entry directory \\
\hline
Provenance + integrity &
Workflow logs, optional per-file hashes, manifest checksums, and metadata used to generate \texttt{dataset\_fair}
and \texttt{summary}. &
Text (YAML/TXT)\\
\hline
Simulation inputs &
Inputs required to interpret the dataset, including configuration files and geometry descriptors referenced by the workflow. &
Text/binary as produced by simulation tools. \\
\hline
\end{tabular}
\end{table}
\fi

\subsection{Verification and Validation (V\&V)}

One of the main drivers for converting the extended MHD NIMROD output within the IMAS framework is the goal of making the Verification and Validation (V\&V) process routine, repeatable, portable, and reusable. In particular, the IMAS framework enables the standardized extraction of equilibria, profiles, and mode-resolved perturbations, which can be used to perform cross-code comparisons and regression testing of the NIMROD post-processing suite as well as the analysis tools. Some examples of representative V\&V targets could be: (i) comparisons of linear growth rates and frequencies, (ii) comparisons of eigenfunction structures, including the mode localization, parity, phase relationships, and spectral content, and (iii) equilibrium/profile consistency checks, such as LCFS geometry, flux coordinates, and profile continuity across the pedestal/SOL interface. All of these could be included in the existing analysis tools in a way that the same scripts can be used for any dataset that conforms to the IMAS schema, minimizing the chance of ``case-by-case'' interpretation and post-processing analysis~\cite{Imbeaux2015}.

\subsubsection{Validation of converted records}
\label{sec:technical-validation}

In this section, we summarize the validation and integrity verifications that are performed to verify the consistency of the IMAS entries deposited after conversion and prior to packaging for downstream analysis.

All of the IDSs follow the same mapping scheme as outlined in the IMAS data dictionary used for this conversion; any necessary transformation of the units occurs when converting NIMROD units into IMAS SI/eV units (Sec~\ref{sec:conversion}). 
The coordinate systems are being enforced by an explicit COCOS 17, as discussed in Sec.~\ref{sec:cocos}, such that the 
sign and normalization of both the equilibrium and derived geometries remain the same for different IDSs.

For unstructured-grid data stored via GGD, the conversion stage verifies:
\begin{itemize}
  \item validity of connectivity arrays and element indexing for the selected connectivity option;
  \item consistency of node coordinates and element definitions across toroidal planes;
  \item absence of NaNs/Infs in stored fields and coordinates;
  \item monotonicity and normalization of 1D profile coordinates (e.g., $\rho_{\mathrm{tor,norm}}$ and
        $\psi_{\mathrm{pol,norm}}$) where the mapping prescribes a one-dimensional data.
\end{itemize}

For the \texttt{edge\_profiles} IDS, where one-dimensional edge-profile coordinates can be geometrically ambiguous, such as multiple disconnected regions sharing the same $\psi_{\mathrm{pol,norm}}$ near the LCFS, the repository provides guidance to use GGD-resolved fields (Sec.~\ref{sec:profiles}).

Each conversion stage records a machine-readable execution trace, including the command lines and input file lists, in the workflow record. Converter interface stores manifest checksums that can be used to detect accidental corruption or unintended modifications, consistent with the provenance strategy described in Sec.~\ref{sec:fair}.

\rev{For the quantitative comparison, the NIMROD and IMAS fields are evaluated on the same poloidal grid. Let $f_i^{\rm dump}$ denote the value obtained from the native NIMROD dump and $f_i^{\rm IMAS}$ the corresponding value reconstructed from the IMAS record at grid point $i$, with $\Delta f_i=f_i^{\rm IMAS}-f_i^{\rm dump}$. For the $N$ grid points at which both values are finite, the relative $L_2$ error and root-mean-square error are defined as
\begin{equation}
		\epsilon_{L_2} =
		\frac{\left(\sum_{i=1}^{N} |\Delta f_i|^2\right)^{1/2}}
		{\left(\sum_{i=1}^{N} |f_i^{\rm dump}|^2\right)^{1/2}},
		\qquad
		{\rm RMSE} =
		\left(\frac{1}{N}\sum_{i=1}^{N} |\Delta f_i|^2\right)^{1/2}.
\end{equation}
The maximum absolute error is $\Delta f_{\max}=\max_i|\Delta f_i|$, while the Pearson correlation coefficient measures the similarity of the spatial variation of the two fields after removal of their respective mean values. Thus, $\epsilon_{L_2}$ measures the error relative to the magnitude of the native field, RMSE and $\Delta f_{\max}$ quantify absolute pointwise discrepancies, and the Pearson coefficient tests preservation of the spatial pattern. The converter-consistent validation includes 4,356 independent field comparisons across 9 time slices, 22 toroidal modes, real and imaginary Fourier components, and a common $361\times641$ poloidal grid, corresponding to approximately $1.008\times10^{9}$ pointwise array-value comparisons. All occurrence-1 fields, including the electron-temperature perturbations used in the Fig.~\ref{fig:vv_contours} comparison, as well as the occurrence-2 ion-temperature and velocity perturbations, are bitwise identical to their corresponding converter-consistent native NIMROD references when read in double precision. The only non-bitwise-identical quantity is the derived ion-pressure perturbation, for which the largest relative $L_2$ error is $9.673\times10^{-17}$, the RMSE is $2.397\times10^{-13}$, and the maximum absolute error is $9.095\times10^{-13}$ in the stored pressure units. The Pearson correlation coefficient exceeds $0.999999999999999$ for all validated field groups. Additional row-level validation results are provided in the companion Data Descriptor~\cite{pankin2026sd}.}

\subsubsection{Validation of physics consistency}
%The conversion process supports validation of physics consistency. This involves comparing derived quantities from the IMAS records against expected behavior. An example of this would be performing reconstruction checks of the toroidal Fourier representations when producing 3D fields using stored harmonics. Another example includes the comparison of flux surface averaged profiles obtained with derived IMAS for the fields stitched on the  $(R,Z)$ grid using binning against the original profiles (Sec.~\ref{sec:profiles}).  Finally, a consistency check is performed for the equilibrium geometry (LCFS and separatrix representation) across the equilibrium and profile IDSs used in coupled workflows.

\begin{figure}[htbp]
  \centering
  \includegraphics[height=10cm]{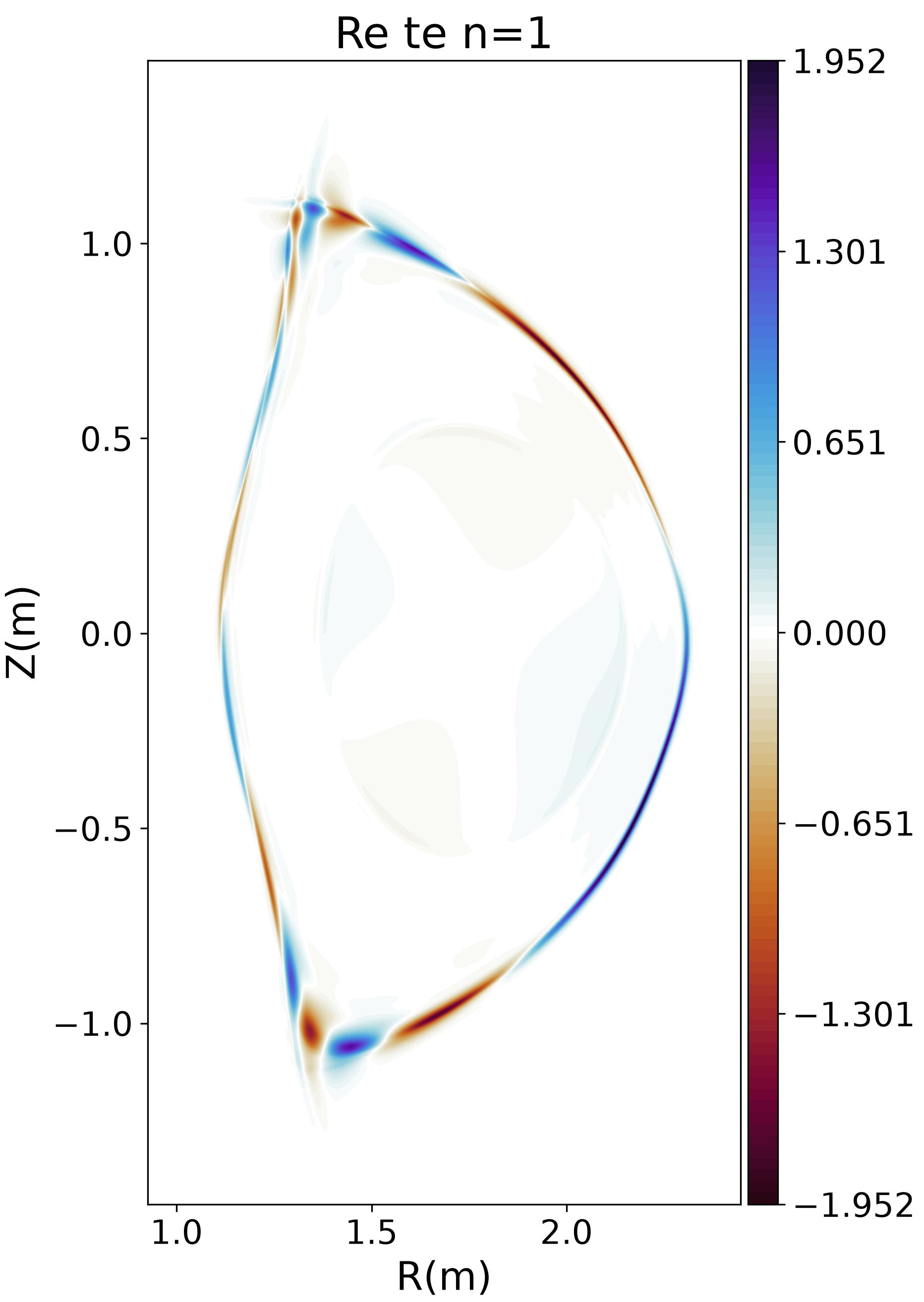}\hfill
  \includegraphics[height=10cm]{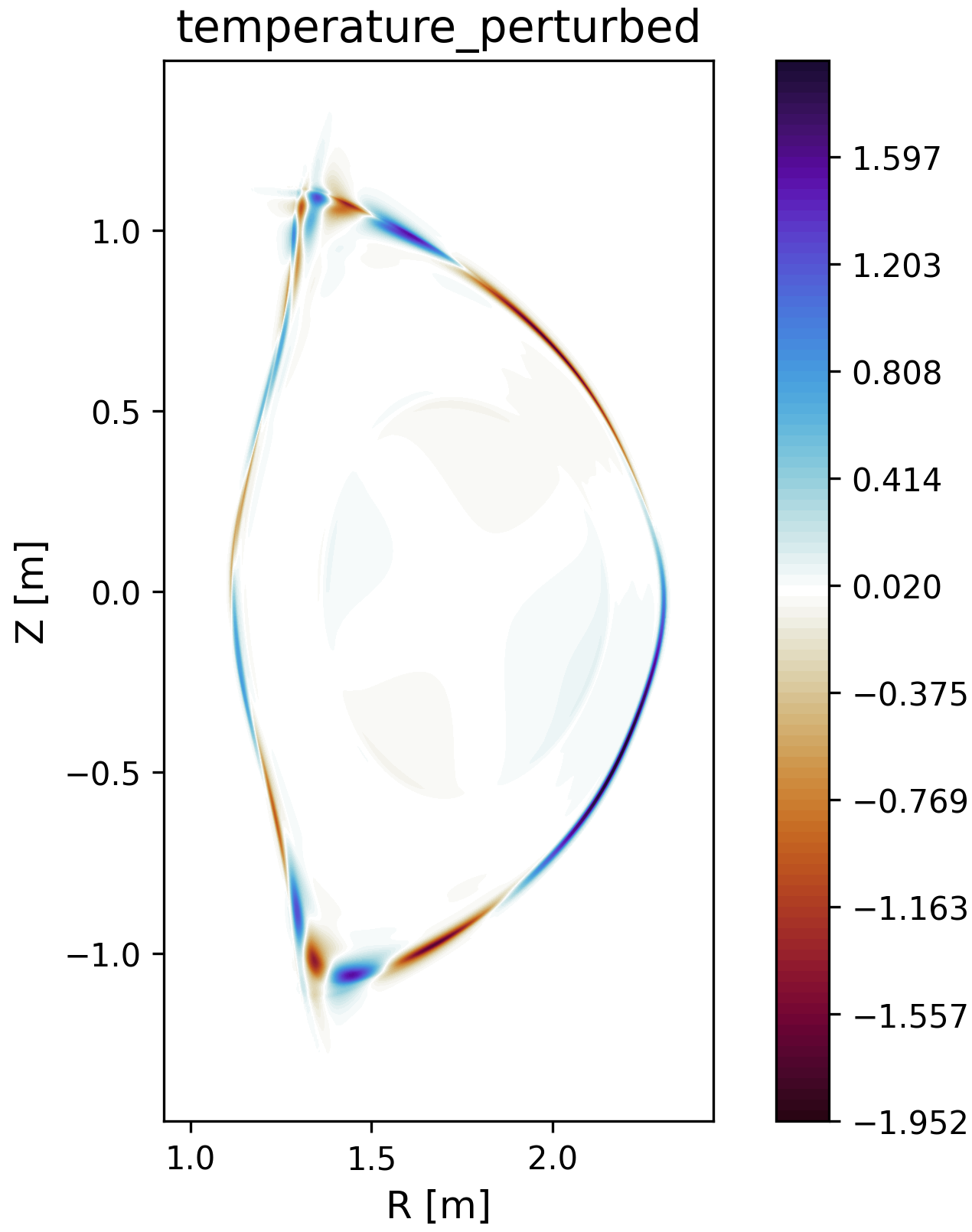}
  \caption{\rev{NIMROD-to-IMAS converter verification using contour reconstruction for an $n=1$ temperature perturbation on the poloidal $(R,Z)$ plane. Left: contour produced directly from the native NIMROD dump using \texttt{nimpy}. Right: contour produced by reading the corresponding field from IMAS using the NIMROD-to-IMAS plotting utility. }}
  \label{fig:vv_contours}
\end{figure}

\begin{figure}[htbp]
  \centering
  \includegraphics[width=0.72\linewidth]{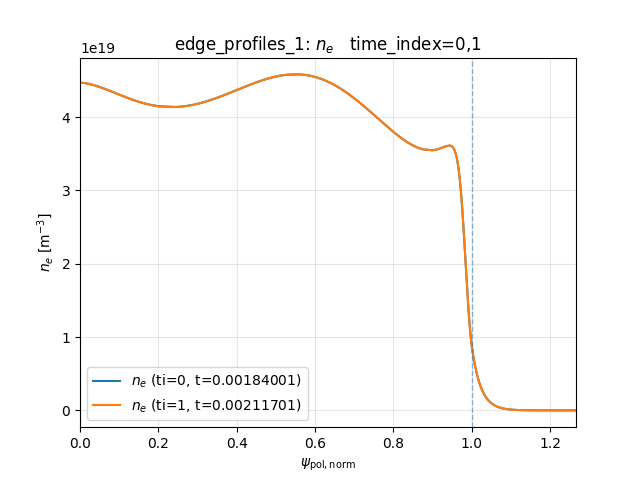}
  \caption{Verification of the internal consistency of one-dimensional profile construction in the \texttt{edge\_profiles} IDS. The electron density $n_e$ is shown as a function of normalized poloidal flux, $\psi_{\mathrm{pol,norm}}$, for two time slices. The dashed vertical line marks $\psi_{\mathrm{pol,norm}}=1$, corresponding to LCFS.}
  \label{fig:edge_profiles}
\end{figure}

In addition to physics-level verification and validation, we also check for the correctness of our NIMROD-to-IMAS implementation by directly comparing spatial contour plots constructed from the original NIMROD dump file and the fields read from IMAS. In Figure \ref{fig:vv_contours}, we show an example for the $n = 1$ perturbation of the electron temperature in the poloidal $(R, Z)$ plane. One contour is constructed from the dump file using NIMROD's built-in \texttt{nimpy} utilities, and the other from fields read from IMAS using a plotting tool included in our NIMROD-to-IMAS conversion package. \rev{The verification is based on the numerical field values and spatial structures rather than on plot shading. The compared features include} its localization, sign changes, amplitudes of the perturbations, and fine-scale features in the vicinity of the edge. This is our end-to-end verification of correctness of our implementation: we are checking that (i) we are reading the correct fields from both codes, (ii) we are not corrupting coordinate information in our data structures and when we write out fields from them, and (iii) we are not losing information about the mode content in our representation of fields in IMAS.

The construction of one-dimensional profiles in the \texttt{edge\_profiles} IDS based on normalized poloidal flux is another internal verification test. The NIMROD dump files don't include flux-surface labels or LCFS location to use as input for the converter. Thus, the converter must establish the $\psi(R,Z)$ function, find $\psi_{\mathrm{LCFS}}$, and bin 2-D fields onto $\psi_{\mathrm{pol,norm}}$ to create these profiles. \rev{Figure~\ref{fig:edge_profiles} compares two time slices as a coordinate-reconstruction consistency test, not as a requirement that two independently evolving plasma states be identical. For the selected slices, the axisymmetric electron-density profile changes only weakly, so the common alignment of the profile features and of the LCFS at $\psi_{\mathrm{pol,norm}}=1$ tests whether repeated reconstruction and binning introduce an artificial time-dependent coordinate shift. Any physical temporal evolution can then be distinguished from a mapping artifact.}

After the fields and associated metadata are stored within the IMAS, regression tests can be defined using compact, code-independent metrics based on queries that are native to the IMAS environment. A first intuitive set of regression tests for linear runs might consist of a set of growth rate and frequency regression tests, where the \texttt{gamma2imas} tool can save the scalars for individual toroidal modes within \texttt{mhd\_linear} and can be set up to mimic the standard workflow for calculating the growth rates, as it is done with NIMROD native \texttt{nimpy} tools. A set of eigenfunction regression metrics might include cross-correlation between the reference and test eigenfunctions, using $L_2$ differences on the same grid, localization of the peak amplitude, and spectral coherence between the stored harmonics of the eigenfunctions.
For equilibrium and profile consistencies in nonlinear simulations, regression tests might verify the invariants of the LCFS and the location of the axis, the monotonicity and normalization of the flux coordinates, spectral properties of the modes and their saturation levels, as well as the consistencies of the derived profiles calculated from the stored equilibrium and core and edge profiles.

The IMAS representation is also very advantageous for multi-code V\&V, \rev{thanks to the common data dictionary used for equilibrium, profiles, and perturbations.} This standardization also allows for comparison with other extended MHD tools and stability codes without the need to create code-specific file translators. In addition, the standardization will help in the validation against experimental measurements, thanks to the processing simplification provided by the capability to analyze simulation results with the same tool sets as is available for the experimental results and IMAS-based synthetic diagnostics. The complete connectivity of GGD is also very important in this context, as it will help in the reproducibility of interpolation and unstructured mesh post-processing and synthetic diagnostics (see Sec.~\ref{sec:ggd_connectivity_options}).

ITER is working to develop a synthetic set of diagnostic forward models, native to IMAS, that use IMAS IDSs as a principal interface. This will provide a direct route for comparison of NIMROD simulations, once they have been converted to IMAS, with experimental diagnostics using a common data model, including machine descriptions, to further support validation and simulation/experiment coupled studies~\cite{Schneider2021,Pinches2025}. In this context, it is desirable to port the Beam Emission Spectroscopy (BES) synthetic diagnostic code developed by the authors~\cite{Pankin2020} to the IMAS interface, allowing the code, together with any code that already utilizes the IMAS infrastructure, such as stability codes, integrated modeling codes, etc., to share the same BES forward model, thereby allowing cross-code verification and validation of the results to be performed. However, this project is outside of the scope of this paper, and it will be considered for future work.

%\section{Applications and Demonstrations\label{sec:sec4}}
\section{Interoperability, schema limitations, and reusable workflows\label{sec:sec4}}
Here, we explore how these converted IMAS files have been used in the real-world reuse of downstream pipelines, and establish what the present day scope and limits of this pipeline are. Our emphasis will be on topics such as interoperability, provenance, and reuse at database level, particularly in relation to machine learning.

\subsection{Coupling to Other Codes and Integrated Modeling Frameworks}
 One of the main benefits of the NIMROD-to-IMAS converter is that the simulation state becomes easily consumable for downstream codes via a common set of IDSs, thus eliminating the need for file translators and reducing the ambiguity of the geometry, normalization, and metadata. In practice, the process of coupling generally requires a well-defined subset of IDSs that represent (i) the magnetic equilibrium and grid geometry, (ii) the background plasma profiles, and (iii) the perturbation content (linear or nonlinear) relevant to the specific applications of MHD analysis. In order to support such an integrated modeling and multi-code approach with a workflow that spans multiple codes, the NIMROD-to-IMAS converter is designed so that the output from the NIMROD code can be directly reused by other codes through a common IMAS data model. The possible applications include:
  
  \begin{itemize}
          \item Standardized inputs and outputs: Using the NIMROD-to-IMAS converter, the NIMROD code can provide the equilibrium geometry and flux coordinates, as well as the background profiles (\texttt{core\_profiles} and, if available, \texttt{edge\_profiles}).   The perturbations and the corresponding scalars for the modes can be provided in the \texttt{mhd\_linear} IDS for Fourier harmonics, eigenfunctions, growth rates/frequencies, which allows the downstream tools to easily compute the total perturbations based on the information provided in the stored spectrum.
  
          \item Interoperability with integrated modeling workflows: The IMAS datasets can be reused to integrated modeling workflows that include transport solvers, stability analysis, and synthetic diagnostics. This is particularly important in view of the development of a synthetic diagnostic infrastructure for ITER based on IMAS. In addition, there is a whole suite of existing integrated modeling tools that already run on IMAS IDS. Expanding this to community diagnostics, such as an IMAS-wrapped BES synthetic diagnostic, allows for forward modeling with multiple simulation codes that share a common interface with IMAS, thereby strengthening verification and validation, as well as validation with experimental data.
  
          \item Coupling scenarios and minimal IDS subsets: An example of a scenario is simulation to diagnostic validation, where a typical example is given by NIMROD $\rightarrow$ IMAS $\rightarrow$ synthetic diagnostics such as Mirnov, BES, and ECE/ECEI, where the minimal IDS subset includes \texttt{equilibrium}, \texttt{core\_profiles}, and \texttt{mhd\_linear}/\texttt{mhd}. Another example is multi-code benchmarking, where one can use IMAS to compare eigenfunction structure and mode scalar data from NIMROD with other MHD/stability codes. Again, in both cases, grid connectivity and full metadata/provenance must be maintained to ensure that any observed differences are due to physics and not to post-processing.
  \end{itemize}

\rev{The same interoperability principle can be used in the reverse direction. %For an experimentally populated IMAS entry, the recommended path is diagnostic IDSs $\rightarrow$ equilibrium/profile reconstruction $\rightarrow$ \texttt{equilibrium} + \texttt{core\_profiles} (and \texttt{edge\_profiles} where SOL information is required) $\rightarrow$ NIMROD initialization. 
The reverse adapter would read the reconstructed equilibrium, wall/LCFS geometry, densities, temperatures, and rotation profiles; convert COCOS=17 quantities back to the NIMROD convention; reconcile units and species definitions; and interpolate the profiles onto the FGNIMEQ/NIMROD mesh. Thus NIMROD can in principle ingest equilibria and profiles generated by other IMAS-capable codes, in the same way that JOREK initialization has used IMAS scenario information~\cite{Massanova2025}. The present release does not yet implement this IMAS-to-NIMROD adapter, because the reverse functionality is already supported with existing tools. However, the complete bidirectional coupling might be considered as a future converter extension.}

\subsection{Federated Campaign Management}
Modern HPC and ML workflows increasingly depend on cross-site, cross-team collaboration, repeated re-analysis, and rapid iteration on quantities of interest (QoIs). Teams are producing different datasets and running analysis on hosts of different resources. Direct download of large production-side output strains network bandwidth, inflates local storage usage, and slows the end-to-end loop from simulation to insight. Federated data management addresses this bottleneck by allowing users to discover and query remote datasets through metadata, then retrieve only the specific subset needed for a given task.

The NIMROD-to-IMAS conversion pipeline provides the semantic foundation for this federation by enforcing an IMAS-consistent data dictionary and harmonizing variable semantics across outputs. In practical terms, the conversion step democratizes the data: once the output is standardized, downstream data-management services can select variables, timesteps, and regions of interest using shared scientific meaning rather than code- or site-specific naming conventions. Campaign management \cite{podhorszki2025hpc} then provides the operational layer for efficient access. It decouples user-facing query workflows from physical storage layout so collaborators can discover and access variables and QoIs across multiple runs and hosts without duplicating full-resolution producer-side data.

Campaign management is a separate toolkit \cite{campaign-code} layered on top of the converted data products. In this layered model, IMAS-to-NIMROD provides schema-standardized scientific data, while campaign services provide metadata-driven federation, location-independent querying, selective remote retrieval, and lifecycle-aware data orchestration. The campaign archive (.aca file) stores dataset and file catalog records and location/provenance context, while primary variables remain in native ADIOS2 or HDF5 files. Query execution follows a two-stage path: metadata resolution (identify dataset, variable, and location) followed by selective data access (retrieve only the relevant slices/blocks based on timestep, coordinates, statistics, etc.). For remote access, the workflow can use \texttt{adios2\_remote\_server}, launched on the data-resident host to execute read operations and return only requested data slices to the client-side ADIOS program. In addition to selective extraction, campaign-oriented workflows can integrate data-reduction actions, including downsampling and lossy compression of extracted payloads before transfer, further reducing data movement and accelerating delivery to users. Possible applications with federated campaign management include:
\begin{itemize}
    \item Remote access and partial data query: In a demonstration, the original NIMROD dump files has size of about 1.38 GB. After NIMROD-to-IMAS conversion, a campaign archive is constructed to register the dataset's metadata and storage locations at the OLCF Frontier. This campaign archive file (e.g., \texttt{xx.aca}) is a lightweight, portable metadata handle, which can be shared among project participants and synchronized to their local or cloud computing devices. If a participant wants to look at a $32\times32$ 2D block of the \texttt{equilibrium.time\_slice.profiles\_2d.0.psi} variable at a designated location, instead of moving the entire dataset, an \texttt{adios2\_remote\_server} will be launched via SSH, and only 8 KB of raw data will be transferred from OLCF to the user's local server.
    \item Local visualization with reduced data: For users who do not have access to the computing nodes at the remote site, \rev{running visualization jobs directly on a remote host can be challenging.} The alternative approach is to download the region-of-interest and launch the visualization locally. This task will require a reader that can process variable block-by-block. This is exactly how the campaign management process ADIOS-formatted \cite{godoy2020adios} data in general. By saving the NIMROD-to-IMAS converted data in ADIOS's bp format, a campaign script can be created to call ADIOS Block Selection to download only blocks of interest. Since Paraview works with ADIOS's block-based data structure, as the examples in \cite{podhorszki2025hpc} illustrated, a sparsely filled global space can be directly visualized using Paraview. 
\end{itemize}

\subsection{Relevance to AI/ML database integration}
Large simulation campaigns are now often treated as data production campaigns, with ensembles of simulations run on multiple machines, physics codes, mesh/resolution parameters, and parameter scans, and subsequently reused for verification/validation, synthetic diagnostics, and AI/ML modeling. A significant bottleneck in the process remains the fact that many MHD and stability codes output their results in code-native formats, making harmonization of runs and long-term reuse expensive and prone to errors as the simulation campaign grows to thousands of runs.
  
The original motivation behind the development of the IMAS data model, based on integrated modeling and the representation of experimental and simulation data in a form that allows for interoperability between them and aligns very naturally to the requirements of ``ML-ready'' databases.
  
In the NIMROD-to-IMAS interoperability workflow that is presented in this paper, a key insight is that IMAS is not simply a file format, but a data model that includes a metadata/provenance layer: the physics content lives in equilibrium/profiles/MHD IDSs, but ‘meta IDSs’ like \texttt{summary}, \texttt{dataset\_fair}, and \texttt{workflow} provide discoverability, enable database-wide filtering, and provide a record of how an artifact is created.

From an AI/ML enablement standpoint, the following characteristics of IMAS significantly reduce dataset friction compared to ad-hoc campaign datasets. First, the distinction between physics-focused IDSs such as \texttt{equilibrium}, \texttt{core\_profiles}, \texttt{edge\_profiles}, \texttt{mhd}, and \texttt{mhd\_linear}, and database-focused ``meta IDSs'' such as those related to database fields and database tags facilitates a tiered access model where fast scan and filter operations can precede slower array retrieval. This tiered access model aligns with the typical ML workflow, where inexpensive dataset indexing precedes expensive tensor operations. Second, IMAS offers a structured format for dense arrays through the GGD format, including support for unstructured mesh coordinate encoding and optional connectivity. For MHD datasets on unstructured finite element meshes, this is important as it removes any ambiguity regarding resampling and neighborhood definitions that can lead to ``silent'' dataset drift as different teams post-process the same dataset with different resampling or neighborhood definitions. Third, IMAS encourages self-describing datasets, and the NIMROD-to-IMAS workflow explicitly uses this by including fields such as code name and version, run IDs, conventions, and provenance for the conversion pipeline. This directly supports reproducibility and can help identify distribution shifts during retraining, such as retraining on a mix of old and new versions of the converter or on a mix of runs with different physics options.

During the systematic labeling of data, the IMAS stores the following information: the stability status (stable or unstable), dominant mode structure,  growth rates, localization, saturation amplitudes, and time-to-event markers. An additional focus of the IMAS, that applies specifically to preparing the data for AI/ML applications, is the management of labels and features, which can be treated as versioned data products as parts of the scientific record after their usage in training AI/ML models.
  
The conversion of data from NIMROD into IMAS follows a multi-stage conversion pipeline, which includes input mappings, dump conversions, and post-processed stability information. The provenance information on workflow steps as described in Sec.~\ref{sec:provenance} describes the origin of the data is especially relevant in defining the data label, including its definitions and NIMROD-to-IMAS scripts version numbers used to create it. Thresholds, time windows, and interpolation/grid definitions are equally important when converting the 3D mesh fields into scalar values. Through this process, a total of three levels of features can be recognized, all of which will be defined as data products using repeatable IMAS native queries:

\begin{itemize}
  \item Tier A: Lightweight scalar data descriptors. Examples of this dataset type would include the equilibrium properties, scalars, and non-dimensional properties, as well as reduced stability output (or pedestal) property type descriptors. This property type is easy to compute and can provide a large amount of information about data rapidly for exploratory purposes.
  
  \item Tier B: Structured but compact representations. Examples of these representations of structured datasets are represented by radial profiles on a 
  standardized flux grid or reduced-dimensional representations of eigenfunctions, or spectral representations of perturbations (for example, kinetic and magnetic energies as functions of mode number), and data that has been consistently derived from the GGD's neighborhood statistics. This tier supports hybrid models that can work with both scalar and structural representations.
  
  \item Tier C: Field-level representations that include the field-level mesh-based representations stored in GGDs. Such field-level representations can be used for graph-based learning and operator-learning models. These models depend on high-fidelity field representation and geometry, where coordinates and connectivity are necessary. In Sec.~\ref{sec:ggd_connectivity_options}, we discussed how the geometry representation depends on the resampling techniques. In particular, the \texttt{hex} GGDs are convenient for the tensor types NNs, but they could potentially be problematic for ELM or EHO simulations in the locations close to the separatrix due to the necessity to interpolate and risks not preserving MHD structures during these interpolations.
\end{itemize}  
  
An IMAS-based campaign dataset can become an ``AI/ML dataset'' if it satisfies specific requirements on metadata, performance, and governance.

\subsubsection{Metadata requirements\label{sec:meta}}
At a minimum, the datasets need to be self-describing enough so that some external process or user can answer some fundamental questions without needing any out-of-band knowledge~\cite{Wilkinson2016}.

The first requirement is to include an identifier for the tokamak name, discharge number, time ranges, and run ID, along with a set of associated scenario tags. The second thing to point out is that for large-scale operation, the emphasis on compact access via IMAS’s summary-style access allows for fast indexing and filtering for the purpose of scaling up operations.

Data identification and other important information about the dataset also need to provide information about the numerical configuration and physics model that have been used in order to generate the data. This requires a description of the following: the code name and version number used; the configuration parameters that were set for the run; and the coordinate/normalization schemes applied to the dataset in order to facilitate the physical interpretation of the data.

Workflow records allow for transparency in where items originated and how they are processed. The continued accumulation of workflow records, rather than replacing the previous record, means that there will always be an `unchanged' version of the previous work, preventing unintentional alterations to the dataset. Furthermore, the \texttt{dataset\_fair}, which is an IDS container for holding important metadata associated with the dataset, provides access to the generated dataset using machine-learning-governed data as well as to metadata associated with that generated dataset, and produces information needed for reproducibility of any dataset created.

The data should have distinct representations of coordinates as well as distinct representations of SOL and confined areas. Maintaining the resolution of meshes that include information on the resolution metadata associated with that mesh will be significant to resolving uncertainties in the representation of 1D edges for cases where there is more than one connection of the regions. At this time, the information on mesh resolution is not part of IMAS metadata IDSs such as \texttt{summary} or \texttt{dataset\_fair}.

A minimum set of metadata must be developed and adhered to prior to any run being added to the ML-ready database. The four pillars of these data contracts are the specific branches and fields of the IDS that need to be defined; the requirement for enumerations and tags when classifying runs/scenarios; the requirement for the use of provenance information (e.g., code version, converter version, config hashes); and the requirement to perform integrity checks (e.g., units of measure, coordinate monotonicity, and array shape) during the development of the NIMROD-to-IMAS converter.

\subsubsection{Performance requirements\label{sec:perform}}
A dataset may be semantically correct but not usable due to the slow ingestion and querying speeds of the data for ML applications. Datasets also need to comply with performance requirements including the following: (1) Datasets may be very large, with size measured in multiple terabytes, for example, non-linear simulations that are dumped frequently and with very high fidelity that have been created for the purpose of interpretation; and (2) The database format and any conversions used to create it must not suffer from ``small writes'' and should have the chunk-based I/O path, as well as batching, discussed previously in this paper in relation to the design of converters.
  
In regard to ML data loading, when there are frequent needs to access specific fields, such as profiles and scalars, on many runs, and accessibility/availability of compact Summary Identification Schemes to filter fields without having to scan large arrays becomes important.
  
We already discussed the problems associated with IMAS's hierarchical levels and verbose metadata, causing large file sizes and not-optimized I/O issues to the GGD connectivity information. This should also be considered a problem in the AI/ML-enabled application of IMAS and recorded as a formal requirement. For example, clearly establishing which dense artifacts are preserved at full fidelity (the original-file-size reference), which artifacts will be downsized (lower than the original file size), and which artifacts can be produced on demand via the provenance information associated with the related scripts.
  
A major element of successfully implementing an archive-based approach for performance strategies is storing original physics data (known as canonical), derived data (a lighter-weight version of what was stored), and derived quantities (derived from other data) in a format that can reliably be recomputed if necessary (for example, if a definition or label changes). The use of a workflow-record framework in the dataset provides a consistent approach towards meeting these goals.
  
\subsubsection{Governance requirements that prevent ML pitfalls\label{sec:gov}}
There are many aspects of ML pitfalls that use the simulation databases. These aspects include not only model structure and integrity of simulated physics, but also how the various data sets are governed. Pitfalls associated with ML approaches related to data governance include but are not limited to data leakage between data sets, hidden changes in labeling and feature definitions during a campaign, blending results from incompatible software or configuration assumptions, and unknowns related to out-of-distribution data sets making the results appear to be more doubtful than they actually are. The pitfalls associated with the IMAS simulation generated campaign data sets are particularly relevant due to the potential differences in outputs through the use of different converters, post-processing methods, interpolation methods, and labeling methodologies. Therefore, it is critical that data governance provide adequate assurance that observed differences in model performance are attributed to either differences in simulation physics or differences in the data production process.

The conversion processes that use conversion scripts or any additional post-processing scripts that include the feature definition logic are expected to have their own versioning system. This information should be included as a part of the scientific record for every run and every derived product. All this information falls under the requirement of having a data management scheme that is centrally controlled. A large number of expected failures in ML development is related to data lineage and not to the underlying surrogate models. An example of this would be the perception that the accuracy of the model has changed; however, in actuality, the model has not changed due to changes in physical attributes of the plasma, but rather due to some previous issues such as a change in the converter used, an alternate set of thresholds being used to define a stability label or a change in the interpolation rules being used to generate a feature from a mesh. Semantic versioning can be used to differentiate between backward-compatible bug fixes versus new features being added and to differentiate between definition-affecting changes. This is to say that data lineage can be considered as a dynamic governance tool instead of being just a passive record of the components used in the specimens or the processes by which they were created. Consequently, provenance allows for reproducibility, assists with controlled comparisons of results between several campaigns, and protects against wrong interpretations of how plasma behaved due to variations in how the data has been processed.

Another approach to address the governance requirements is to split the training, validation, and test datasets based on the relevant physical keys used in the experiment, such as the scenario class, the parameter scan ID, the equilibrium family, and the time range, rather than randomly selecting data frames. Doing this prevents contamination from near-duplicate runs or subsequent time slices. This procedure is also consistent with the general recommendation of using filtered metadata for reproducibility.

Beyond just having FAIR metadata, the machine learning community has been increasingly discussing the need for some sort of formalized document describing datasets. These documents can be referred to as ``datasheets'' or ``data cards''~\cite{Mitchell2019,Gebru2021} and provide information regarding the intended use of a dataset, including the potential limitations and any biases within it. For fusion simulation database datasets, this includes documentation pertaining to the physical model's range of validity, numerical resolution ranges, and any area identified as being outside of the distribution or training set, such as a boundary condition or a unique mix of impurities. 

An IMAS-based database intended for AI/ML use should enforce a version policy, a leakage-free data-split policy, and dataset-level documentation that makes the intended use, limits of validity, and known biases of the resulting data products explicit.

\subsubsection{Satisfying metadata, performance, and governance requirements in the NIMROD-to-IMAS workflow}
The NIMROD-to-IMAS conversion workflow is structured so that the metadata, performance, and governance requirements defined in Sections~\ref{sec:meta}-~\ref{sec:gov} are satisfied. In fact, the conversion tools are implemented as a staged workflow with three principal components. The first, \texttt{input2imas}, converts from primary simulation inputs, including equilibrium data file, \texttt{geqdsk}, and kinetic profiles in peqdsk and corresponding namelists into the IMAS-specific representation of equilibria and profiles for a tokamak with optional wall information. The second, \texttt{dump2imas}, converts from both linear/non-linear NIMROD state dump(s) to IMAS time-varying equilibria/profiles mode-resolved perturbation models (\texttt{mhd\_linear}) and GGD models \rev{for mesh-resolved three-dimensional field snapshots (\texttt{mhd}) at selected} times. The third, \texttt{gamma2imas} extracts and saves from NIMROD mode values, such as growth rate and optionally frequencies, directly to \texttt{mhd\_linear} IDS for different toroidal mode numbers.
  
Both physics-related IDS content and metadata that reflect the dataset are part of all converters. This gives a downstream user or an automated pipeline the ability to determine what the dataset is and how to interpret its contents without having external knowledge of the dataset. Absent from run identity are the shared conventions of pulse and run, occurrence conventions, machine/scenario descriptors, and dataset-level annotations; however, they are preserved in the IMAS-entry organization and compact containers for metadata. In addition, configuration provenance associated with the conversion process is maintained in several forms; key inputs are also retained as IMAS-native physics quantities (such as the equilibrium), while original NIMROD-relevant namelists are preserved as XML strings in the IMAS \texttt{code.parameters} field. This provides a self-contained reconstruction of model and number options.
  
The conversion tools have been designed for large-scale operations. They will minimize the unnecessary amplification of rewrites as time slices are appended and support the selective persistence of dense (versus derived) quantities as well. Specifically, appending of time slices will occur incrementally where supported by the accessing layer and the mesh/connectivity payloads required to represent the GGD will be written in a backend-consistent format to minimize overhead when loading large ensembles of GGD data to speed up the ingestion and loading of real-world application patterns, such as when filters are used to load only candidate runs after loading dense array-like data elements or data elements that can be represented as dense arrays.
    
At each conversion stage, workflow command is stored as metadata information including the versions and repository names, execution commands and, optionally, input artifacts checksums. This detailed provenance data from the conversion process enables (i) rigorous reproducibility of datasets created; (ii) auditing definitions of labels/features used in connection with software used to create those definitions; and (iii) diagnosis of any distributional shifts seen \rev{when machine learning models are trained on multiple campaigns}; and can be useful for determining why a model’s performance has degraded against historical benchmarks.

\section{\rev{Summary and future work}}

The research presented in this paper provides two main outcomes. The first outcome is the development of a realistic methodology to convert NIMROD's input/output data into IMAS-compliant data, allowing for the reuse of such data in future validation studies, coupling studies, and post-processing applications. The second outcome is the identification of a list of lessons learned from this process, which includes \rev{the gaps within the IMAS framework for extended magnetohydrodynamics data} and the requirements in terms of metadata, provenance, and performance for successful downstream AI/ML analysis.

The developed workflow has applications for AI and ML workflows with scientific interpretability and employs a conversion toolchain that stores equilibrium, flux-labeled profiles, and mode-resolved perturbation fields in standardized IDS containers. This workflow also provides dataset-level provenance and metadata compliant with FAIR principles. The provenance and metadata are associated with the IMAS native data and workflow descriptors. This approach supports reproducible labeling, feature extraction across large datasets of simulation results, and provides identification of schema and interpretation gaps that require near-term modifications to address deficiencies regarding: (i) absence of canonical locations in IMAS schema for dissipation profiles for resistivity, conductivity, and viscosity, (ii) MHD simulation with multiple species require use of multiple occurrences of \texttt{mhd\_linear} IDS, and (iii) ambiguity in the edge profile interpretation for one-dimensional plasma profiles derived from MHD simulations. These observations do not make the IMAS schema amenable to durable ML-ready containers, but they are intended to improve interoperability between IMAS datasets produced for MHD simulations.

The main result is an end-to-end NIMROD-to-IMAS conversion workflow. All NIMROD inputs and most outputs are mapped to standardized physics containers. Inputs related to the equilibrium and 1D kinetic profiles are stored in standard \texttt{equilibrium} and \texttt{core\_profiles} IDSs. The 1D profiles in the \texttt{core\_profiles} and \texttt{edge\_profiles} IDSs, which are used for storing flux-averaged plasma profiles, can be used for downstream access by other codes in the integrated modeling frameworks. These IDSs are well developed and stable. As long as the potential customers of workflow recognize that the 1D profiles in the \texttt{core\_profiles} IDS are stored as a function of toroidal flux and the 1D profiles in the \texttt{edge\_profiles} IDS are stored as a function of the normalized poloidal flux (ambiguity mentioned previously), the downsteam access is enabled using stable IDS semantics and coordinate definitions. \rev{The workflow stores flux-averaged one-dimensional profiles in \texttt{core\_profiles}/\texttt{edge\_profiles}, mode-resolved Fourier perturbations in \texttt{mhd\_linear}, optional mesh-resolved edge quantities in the \texttt{edge\_profiles} GGD branches where applicable, and reconstructed full 3D fields in \texttt{mhd}.} The \texttt{mhd\_linear} IDS also includes records on mode stability, including the growth rates and optional frequencies. This information is important for the development of reduced-order ML representation. The database results include the physics information and other structured metadata along with the data provenance; workflow steps with options and dataset descriptors are stored using IMAS-native \texttt{workflow} and \texttt{dataset\_fair} IDSs. This makes the NIMROD-to-IMAS workflow consistent with the FAIR principles for data auditability and reproducible rebuilds across simulation campaigns. Portability of the IMAS data is ensured through IMAS-Python data management approaches with standardized backends such as HDF5 and netcdf with their associated metadata fields that hold the Data Dictionary and conversion script version information.

One of the key strengths of this approach is that it allows for semantic harmonization of extended MHD data for profiles, equilibria, and perturbations since this information is exported from ensembles using the same IDS scheme. This allows for stable labeling and feature-extraction definitions for all parameter scans across different datasets — even as post-processing software evolves over time. The second benefit is that governance is enforced via construction through the schema of the \texttt{workflow} and \texttt{dataset\_fair}. These schemas define the tools' identities, the stages of the pipeline history, identifiers for all datasets, and any reuse constraints that enable provenance-based attributes to retrain and diagnose distribution shifts. The third strength is practical scalability: the workflow is designed around the analysis of conversion scripts for appending time slices and optionally keeping dense unstructured grid representations of entire datasets only when required. This allows matching ML requirements and patterns for metadata-filtering or selecting tensor access.

There are also several limitations of the workflow. In particular, IMAS does not provide a well-defined IDS to store the dissipation profiles for MHD simulations. The absence of unambiguous dissipation profiles in IMAS means that data must be reconstructed from stored profiles and the provenance of the code input. Although the reconstruction is possible, this generates additional work in configuring and managing these workflows to support training feature extraction if full dissipation profiles are required. Ambiguity in storing the 1D edge profiles as functions of normalized poloidal flux in the topologically complex region near the separatrix requires proper documentation and description. In addition, the performance portability remains a challenge: very deep hierarchical structures and large arrays of structures (AoS) can place a load on specific backends/access layer stacks. Here again, we developed specific means to mitigate this load by balancing the IMAS-Python conventions with backend-specific optimizations. More complete benchmarking across storage formats and HPC filesystems is still needed, and it will be a part of future work.

The NIMROD-to-IMAS converter allows creation of datasets free of data leakage through the use of explicit metadata filters with run identity, scenario tags, and provenance fields stored in the stable IMAS schema. The stability of labels (in terms of supervised targets such as stability outcomes, dominant modes and growth rate ranges) is preserved, since they will all have used a consistent set of IDS definitions throughout the duration of the simulation campaign. Additionally, to avoid or detect any possible bias, any changes to converters or label logic will be traceable through the provenance information associated with the workflow. The design of the dataset allows for it to be utilized in numerous ML applications, including: (i) use of surrogate models for establishing the connection between either equilibria and MHD stability; (ii) quantifying uncertainty by way of propagating the uncertainty present in input profiles and profiles used for analysis; and (iii) transfer learning in the cases where the experimental and simulation data are created using a hybrid form using the same IMAS data models that share common IDSs and coordinate systems. The absence of dissipation profiles does not affect the reproducibility of ML runs. Resistivity and other dissipation profiles can be reconstructed from temperature and density profiles, and input parameters using specific definitions such as \rev{the Spitzer formulation for resistivity}. Sufficient documentation of the assumptions can be stored using the provenance information. The requirement to reconstruct the dissipation profiles adds to the complexity of the analysis pipeline and it establishes the need for changes in the IMAS schema by providing an integration of the dissipation profiles for both \texttt{mhd} and \texttt{mhd\_linear} IDSs. These enhancements to the IDSs will provide a greater level of interoperability between various code implementations and give a common basis for comparison between datasets developed from ML algorithms across multiple extended MHD modelling groups. Other potential extensions to the \texttt{mhd\_linear} IDS are support for multi-species MHD simulations. However, despite all of the limitations present in the IMAS schema, we are developing practical alternatives using the present IMAS functionality with appropriate provenance documentation.

Expanding IMAS capabilities to attain restart-ready completeness of extended MHD workflows is an important long-range objective. Accomplishing this will require additional schema capabilities for dissipation functions, multicomponent state quantities, and some diagnostics that have not yet been addressed.

In this research, we show that we can use the current IMAS schema as a solid basis for an extended MHD database foundation for ML applications. The database structure provides sufficient support for ML by using existing IMAS schemas and combining with existing physics content using the existing IDS semantics, and meeting the FAIR standards. This includes providing data provenance and dataset metadata in a FAIR manner; both of these elements help create reproducible labels and features for all datasets that adhere to the current IMAS schema.

%\section{Data Records}
%\label{sec:data-records}
%\subsection{Known limitations}
%Limitations of the data model are known, and workarounds have been implemented, such as for missing fields for multiple ion species. Some of these limitations are discussed in Sec.~\ref{sec:gaps}. As discussed, we do not store 2D dissipation profiles in the dataset, because these no obvious place to store these profiles in the current IMAS DD schema, but they can be reconstructed with the information that is already available in the IMAS dataset.

\section*{Data and Code Availability}
The NIMROD-to-IMAS conversion workflow described in this Article uses the Python tools \texttt{input2imas.py}, \texttt{dump2imas.py}, and \texttt{gamma2imas.py}. The code used to generate the deposited IMAS entries, including version information and example commands, is archived at \emph{https://github.com/PrincetonUniversity/nimrod2imas} under MIT license.

The IMAS Python interface provides access to the datasets in the repository for IMAS DD v4.1 at the time of deposit. The README specifies the IMAS major version that is tested, along with the minimum version of Python required in order to generate the archives and the necessary IMAS bindings. 

All datasets associated with this work are deposited in the Princeton Data Commons (PDC) repository that is accessible using this URS: \indent\url{https://datacommons.princeton.edu/discovery/catalog/doi-10-34770-z037-2b57}. The repository provides a persistent identifier DOI 10.34770/z037-2b57 for the dataset. The data are released under the Creative Commons Attribution 4.0 International license (CC BY 4.0).\footnote{\url{https://creativecommons.org/licenses/by/4.0/}}

A detailed description of the deposited IMAS entry, including the contents of individual IDS files, dataset size, validation metrics, and reuse notes, is provided in a companion Scientific Data Data Descriptor~\cite{pankin2026sd}.

\section*{Funding and Acknowledgments}
The research described in this paper was conducted at Princeton Plasma Physics Laboratory, a national laboratory operated by Princeton University for the United States Department of Energy, (Center for Edge of Tokamak Optimization) CETOP SciDAC-5 project, under Prime Contract No. DE-AC02-09CH11466. The United States Government retains a non-exclusive, paid-up, irrevocable, worldwide license to publish or reproduce the published form of this manuscript, or allow others to do so, for United States Government purposes. This research used resources of the National Energy Research Scientific Computing Center, a DOE Office of Science User Facility supported by the Office of Science of the U.S. Department of Energy under Contract No. DE-AC02-05CH11231 using NERSC award FES-ERCAP0032316.

The authors thank Dr. Olivier Hoenen from the ITER Organization for careful reading of the manuscript and for providing useful comments and suggestions.

\section*{Declaration of competing interest}
The authors declare no competing interests.

\section*{Author Contributions}
A.Y.P. led the IMAS mapping design and implemented the conversion workflow and validation.
F.E., J.K., A.K., and J.D.-P. contributed to the NIMROD physics use cases and requirements definition. Q.G. contributed to the workflow verification and ADIOS/CAMPAIGN schema development. All authors reviewed and approved the final manuscript.

\printbibliography
\end{document}